\documentclass[aps,prb, print,twocolumn]{revtex4-2}%
\usepackage{amssymb}
\usepackage{amsfonts}
\usepackage{amsmath}
\usepackage{graphicx}
\usepackage{bm}

\begin{document}

\preprint{APS/123-QED}

\title{Wave Analysis and Homogenization of Spatiotemporally Modulated Wire Medium}

\author{Michael Kreiczer and Yakir Hadad*}
\affiliation{The school of Electrical Engineering, Ramat-Aviv, Tel-Aviv University, Israel, 69978}
\email{hadady@eng.tau.ac.il}

\date{\today}

\begin{abstract}
In this paper we develop homogenization theory for spatiotemporally modulated wire medium. We first solve for the modal waves that are supported by this composite medium, we show peculiar properties such as extraordinary waves that propagate at frequencies below the cut-off frequency of the corresponding stationary medium. We explain how these unique solutions give rise to an extreme Fresnel drag that exists already with weak and slow spatiotemporal modulation. Next, we turn to derive the effective material permittivity that corresponds to each of the first few supported modes, and write the average fields and Poynting's vector. Nonlocality, nonreciprocity, and anisotropy due to the spatiotemporal modulation direction, are three inherent properties of this medium, and are clearly seen in the effective material parameters.
As a figure of merit, we also derive the effective permittivity of a plasma medium with spatiotemporally modulated plasma frequency. This comparison is interesting since the plasma medium can be considered as the effective medium that is obtained by a stationary wire medium. We validate that homogenization and spatiotemporal variation are not necessarily interchangeable operations. And indeed, in certain parameter regimes the homogenization should be performed directly on spatiotemporally modulated composite medium, rather than first homogenize the stationary medium and then phenomenologically introduce the effect of the space-time modulation.
\end{abstract}

\maketitle

\section{Introduction}
The study electromagnetic wave dynamics in time-varying media  goes back several decades ago \cite{Transmission of electromagnetic waves into time-varying media, Velocity modulation of electromagnetic waves}. Recently, there is a new blooming of research in this direction especially in the context of new developments toward the next generation in the research of metamaterials. Such a progress is accompanied by new-technological abilities that make the actual realization of engineered time-varying media plausible.  The benefits from removing the restriction caused by time invariance, and adding another dimension for engineering - the time, has opened a whole new area of technological possibilities.  Linear time variant (LTV) metamaterials have been recently used for various applications, such as: inverse prism, frequency conversions, temporal band gap, magnetless nonreciprocity, time reversal and effective permittivity realization by temporal switching \cite{Inverse prism based on temporal discontinuity and spatial dispersion,Changing the colour of light in a silicon resonator,Nonreciprocal Thermal Material by Spatiotemporal Modulation,Modulated phononic crystals: Non-reciprocal wave propagation and Willis materials,Magnetic-free non-reciprocity and isolation based on parametrically modulated coupled-resonator loops,Reflection and transmission of a wave incident on a slab with a time-periodic dielectric function,Temporal photonic crystals with modulations of both permittivity and permeability,Time reversal and holography with spacetime transformations,Optical time reversal from time-dependent Epsilon-Near-Zero media,Wave propagation in periodic temporal slabs. In: 2015 9th European Conference on Antennas and Propagation,Temporal photonic crystals: Causality versus periodicity,Time reversal of electromagnetic waves,Effective medium concept in temporal metamaterials,A Generalization of the Kramers-Kronig Relations for Linear Time-Varying Media, Hadad2015, Hadad2016, HadadSounas2019}. Moreover, since many of the known physical bounds on wave engineering are based on the LTI assumption, by enabling the wave network to be time-varying  we may expect overcoming these physical LTI bounds  \cite{Foundations for Microwave Engineering,Electromagnetic Metamaterials: Past Present and Future}. See for example  \cite{Beyond the Bode-Fano Bound: Wideband Impedance Matching for Short Pulses Using Temporal Switching of Transmission-Line Parameters} for the Bode-Fano bound for impedance matching for LTV microwave circuit, and \cite{Temporal Switching to Extend the Bandwidth of Thin Absorbers} for overcoming the bandwidth bound in matching small antennas.


On a parallel route. Effective medium theory which is classical in electromagnetics and dates back to the 1904 work of Maxwell Garnett \cite{Colours in metal glasses and in metallic films} has become a cornerstone in the context of metamaterials. It is an essential tool to model and analyze effective media with various types of meta-atoms \cite{Plasma imulation by artificial dielectrics and parallel plate media,A Study of Artificial Dielectrics}. The effective medium approach, also termed homogenization, allows us to transform from the composite material, and its microscopic structure to a new  equivalent material having, macroscopic, smoothed properties.
A particularly interesting example of effective medium is the wire medium of its various types.
In its simplest form, the wire medium consists of a two dimensional lattice of infinitely long perfectly conducting wires surrounded by an homogeneous dielectric material.
Despite its apparent simplicity that enables a relatively easy to follow analytical modelling, interestingly, the wire medium gives rise to various peculiar properties such as strong anisotropy, effective plasma-like dispersion, as well as strong-nonlocality \cite{Nonlocal permittivity from a quasistatic model for a class of wire media,Dispersion and Reflection Properties of Artificial Media Formed By Regular Lattices of Ideally Conducting Wires,Photonic bad gaps for array of perfectly conducting  cylinders,Green's function and lattice sums for electromagnetic scattering by a square array of cylinders,Photonic band structures of two dimensional systems containing metallic components,Photonic band structures of two dimentional systems fabricated from rods of a cubic polar crystal,Photonic bands of metallic systems I. Principle of calculation and accuracy,Metallic Photonic band gap materials,Radiation from elementary sources in a uniaxial wire medium}. Various extensions to the basic wire medium have been studied, including loaded-wire medium that is  relatively straightforward to analyze with various loading types \cite{Tretyakov}.

In this paper we study electromagnetic waves that are propagating in a capacitively-loaded wire medium that  undergo a spatiotemporal modulation. See illustration in Fig.~\ref{Fig1}. Our goal in this work is two fold.
First, taking into account the complete interaction between the spatiotemporally modulated loaded wires, we explore the various wave phenomena that can be supported by this medium. Specifically, we study ($i$)  non-reciprocal and anisotropic propagation due to the spatiotemporal modulation, ($ii$) the existence of extraordinary waves that propagate at frequencies that are below the so-called plasma frequency of the stationary wire medium (that is, in the absence of modulation), even with negligible modulation index, and thus are akin to the Whistler mode in magnetized plasma \cite{Ishimaru}, and consequently ($iii$) the emergence of  Fresnel drag \cite{Fresnel drag in space time modulated metamaterials} already with weak and slow modulation.
The second goal, is to derive effective medium properties when the meta-atoms (the loaded wires) are time-modulated. In particular we explore the effect of taking into account the time-modulation already in the homogenization process, as opposed to first homogenize the stationary medium and later  introducing the time-modulation into the effective properties. In other words, we try to address what is the effect of the spatiotemporal modulation on the homogenization. Related to that, is the issue of dispersion in the presence of time modulation, this has been  recently  tackled in several works such as \cite{Solis_TV_with_Dispersion, Solis_Generalized_KK}. Here, we provide a  study for the specific wire-medium case using a quasi frequency-domain approach, and take the complete inter-wire interaction that comprises the composite medium, as discussed below.
We pay a particular attention to the wave dynamics near the plasma frequency, there, due to the sensitivity of the zero crossing point of the effective transverse permittivity of the stationary wire medium, interesting phenomena pop up as soon as we introduce weak spatiotemporal modulation.


The paper is arranged as follow. For the sake of its self-continency,  in Sec.~\ref{Sec_Stationary}, we open the paper with a brief review on stationary capacitively loaded wire media. Then, in Sec.~\ref{Sec Time varying} we introduce spatiotemporal modulation to the capacitive loads and and provide a through mathematical formulation to solve the modal problem.  In Sec.~\ref{Sec Near Plasma},  focusing on narrow frequency range near the plasma-frequency of the corresponding stationary lattice, we derive analytic expressions for the dispersion relations of the first lowest order modes supported in the lattice, and study the wave dynamics in this parameters domain. Later, in Sec.~\ref{Sec Homo}, we derive the effective medium parameters of several supported modes, in addition we derive expressions for the averaged electric and magnetic fields, and the corresponding Poynting's vector in this case. Finally, in Sec.~\ref{Sec perturbed cont}, we compare between the wave phenomena in the time-modulated wire medium, and the wave phenomena predicted when the time-modulation is introduced into the effective medium properties of the stationary wire-medium, i.e., after homogenization.
%

\section{Stationary capacitively loaded wire medium}\label{Sec_Stationary}
In the following sections we turn to analyze wave phenomena in spatiotemporally modulated   loaded-wire media. Unavoidably, we will frequently refer to the derivation and the main  results of the corresponding, stationary, case. Therefore, to make the paper self-contained, in this section we briefly overview \emph{stationary}, capacitively loaded wire medium. We mainly follow the  analysis in \cite{Tretyakov} for general wire loading, and stress the more relevant aspects for the spatiotemporal problem to follow.


\begin{figure}[ptb]
\centering
\vspace{-0.0cm}
\includegraphics[width=\columnwidth]{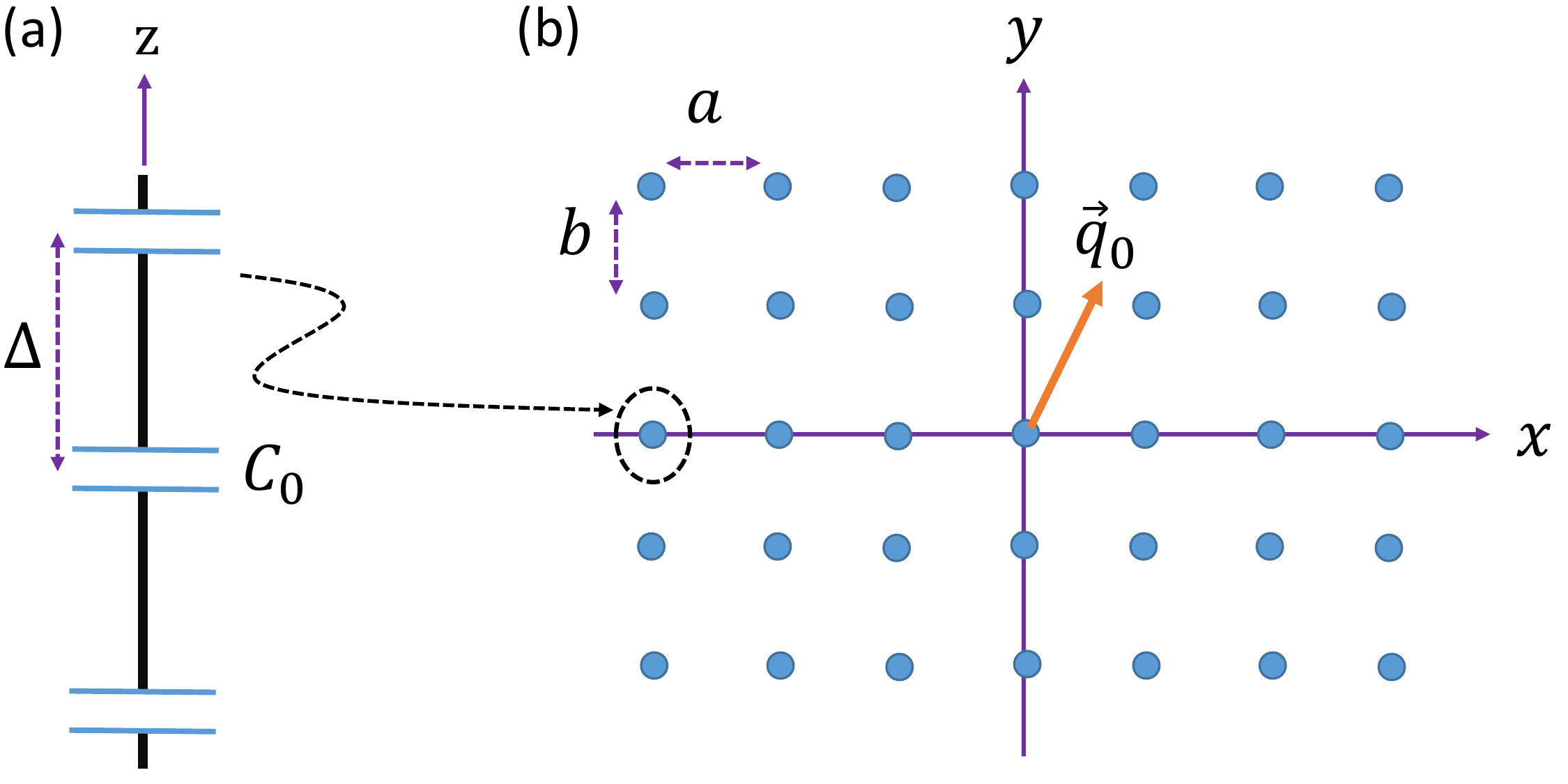}
\vspace{-0.0cm}%
\caption{(a) Capacitively loaded wire. (b) The wire medium, on a rectangular lattice with unit-cell dimensions $a$ along $x$ and $b$ along $y$. In this work we focus on wave that propagate transverse to the wires with $\vec{q}=q_x\hat{x}+q_y\hat{y}$. }
\label{Fig1}%
\end{figure}

Assume that a $\hat{z}$ polarized electromagnetic wave is propagating inside an infinite loaded wire medium as shown in Fig.~\ref{Fig1}. We assume that the wires are surrounded by vacuum with permittivity and permeability $\epsilon_0$ and $\mu_0$, respectively.
The lattice points are given by: ${\vec R}_{m,l} = ma\hat x + lb\hat y$. The induced current on each of the wires is given by $\vec I = \alpha \vec E^{\mbox{\small loc}}$ where $\vec E^{\mbox{\small loc}}$ is the local electric field, namely, the electric field at the wire location but in the absence of the wire itself, and $\alpha$ is the wire\'s susceptibility.
If the wires are periodically loaded by lumped impedance $Z_L$, with periodicity, $\Delta\ll\lambda$, the inverse susceptibility  satisfies \cite{Tretyakov},
\begin{equation} \label{gamma_of_loaded_wire}
{\alpha ^{ - 1}}\left( \omega  \right) =  {\alpha_0^{-1}}\left( \omega  \right) + \frac{Z_L}{\Delta},
\end{equation}
where $ {\alpha_0^{-1}}\left( \omega  \right) = \frac{{\eta k}}{4}H_0^{\left( 2 \right)}\left( {k{r_0}} \right)$. For serial loading of lumped capacitors $Z_L/\Delta=1/(j\omega\tilde{C}_0)$  and $\tilde{C}_0=C_0\Delta$. Here $\eta=120\pi\Omega$, and $k=\omega/c$ are  the free space impedance and wavenumber, respectively, $\omega$ is the radial frequency, $c$ is the speed of light in vacuum, $r_0$ is the wire radius, and $H_0^{(2)}$ denotes the zero order Hankel function of the second type. In light of the  lattice periodicity the Floquet-Bloch wave solution is expected,
\begin{equation} \label{I_m_l_no_perturbation}
{I_{m,l}} = {I_0}{e^{ - j\left( {{q_x}am + {q_y}bl} \right)}}
\end{equation}
where $q_x, q_y$ stand for the transverse (to $z$) components of the wavenumber in the wire medium. Note that throughout this paper we assume for simplicity that $q_z=0$ which is in accord with our initial assumption that the electric field is polarized solely along $z$.
Under the assumptions above, the wavenumber dispersion is found by solving,
\begin{equation} \label{dispersion_cond}
E_z^{\mbox{\small loc}}\left( I_0 \right) = \alpha^{-1}(\omega)I_0
\end{equation}
where $E_z^{\mbox {\small loc}}$ denotes the local field acting on the wire at $m=l=0$. Following \cite{Tretyakov} it is given by,
%
%
\begin{equation}\label{E_local_media}
E_z^{\mbox{\small loc}} = -I_0\frac{\eta k}{2} \left(S_0 + \sum_{m\neq0} S_m e^{-jq_x am} \right),
\end{equation}
with,
\begin{subequations}
\begin{align}\label{S_0_media}
&{S_0} = \sum\limits_{l = 1}^\infty  {H_0^{\left( 2 \right)}\left( {k{R_{0l}}} \right)\cos {q_y}bl} \\\label{S_m_media}
&{S_m} = \frac{1}{2}\sum\limits_{l =  - \infty }^\infty  {H_0^{\left( 2 \right)}\left( {k{R_{ml}}} \right){e^{ - j{q_y}bl}}}.
\end{align}
\end{subequations}
and $R_{ml} = \sqrt{(ma)^2 + (lb)^2}$.
The infinite slowly converging series
in Eqs.~(\ref{S_0_media}) and (\ref{S_m_media}) may be converted
into fast converging series using e.g., Poisson summation \cite{Tretyakov, Field theory of guided waves}.
%
%
For a dense grid, $ka,kb \ll 1$,  the dispersion relation may be approximated by
\begin{equation}\label{dispersion_const}
    \omega\tilde{L}_w - \frac{1}{\omega\tilde{C}_0} = \frac{k\eta}{ab(k^2-q^2)}
\end{equation}
where $q^2 = q_x^2 + q_y^2$ and
\begin{equation}\label{Lwire}
\tilde{L}_w = \frac{\eta}{2c}\left( \frac{1}{\pi}\ln\frac{b}{2\pi r_0}\!+\!\sum_{l=1}^{\infty}\frac{\coth(\pi al/b)-1}{\pi l}\!+\!\frac{a}{6b}\right)
\end{equation}
denotes the per unit length intrinsic wire inductance.
%
%
We note that for the dense grid case, the dispersion equation depends only on $q = \sqrt {q_x^2 + q_y^2}$. Then,  an equation for the effective permittivity along the longitudinal (to $z$) axis is obtained by $\varepsilon_z = \frac{q^2}{k^2}$. This yields
\begin{equation} \label{eps_LTI}
{\varepsilon _z}\left( \omega  \right) = {\varepsilon _0}\left[ {1 - \frac{{k_p^2 {\left( k \right)}}}{{{k^2}}}} \right].
\end{equation}
with
\begin{equation}\label{Plasma_Freq_LTI}
k_p^2(k) = \frac{\eta k}{ab(\omega\tilde{L}_w - 1/\omega\tilde{C}_0)}.
\end{equation}
where $k=\omega/c$, and $k_p$ is termed here and henceforth, the plasma frequency.
%
Nevertheless, note that  unlike the unloaded wire media, for which effectively $\tilde{C_0}\rightarrow \infty$ and consequently $k_p^2 = (\eta/c)/ab\tilde{L}_w = (1/\tilde{C}_b\tilde{L}_w)/c^2$ with $\tilde{C}_b=ab\epsilon_0$ is frequency independent, when the wires are loaded the behaviour is not of a typical plasma since $k_p$ depends on $k$. In the following we shall term   $\tilde{C}_b$ - the per-unit-length background capacitance.
%
%
%


\section{Wire media with spatiotemporally modulated capacitive loading}\label{Sec Time varying}
Our first goal in this paper is to study wave propagation in a spatiotemporally modulated wire media. To that end, we generalize the conventional approach for stationary LTI wire media that was described above. Our analysis is based on the concept of harmonic balance that can be regarded as a \emph{quasi}-frequency domain technique. In this sense, our method is less general then, e.g., \cite{Dipole polarizability of time-varying particles} that suggests, in principle, a direct time-domain venue to that problem. However, since we focus on the particular case of  small time-harmonic capacitance perturbation, we find this methodology, as discussed below, to be more direct and therefore a better fit for this problem.

\subsection{The response of a  single wire}
Assume that the loading capacitance on each of the wires is modulated as,
\begin{equation} \label{capacitors}
C\left( t \right) = {C_0} + \delta C\cos \left( {\Omega t - \varphi } \right)\,\,,\,\,m=\frac{{\delta C}}{{{C_0}}} \ll 1.
\end{equation}
Here and henceforth, $\Omega$ denotes the modulation frequency and $m$ the modulation depth.
In the absence of time modulation, the time-domain counterpart of Eq.~(\ref{dispersion_cond}) is given by $\widehat{{E^{\mbox{\small loc}}}}\left( t \right) = \widehat{\alpha^{-1}}(t) * \widehat{I}\left( t \right)$ where `wide-hat' $\widehat{X}(t)$ represents the time-domain, inverse-Fourier transform, counterpart of $X(\omega)$, and $X$ stands for $E^{\mbox{\small loc}}$, $\alpha^{-1}$, and $I$.    This time-domain perspective enables us to naturally introduce the effect of the time-modulation on the capacitors (see appendix \ref{sec:susceptibility of time variant loaded wire})
\begin{equation} \label{E_time}
\widehat{E^{\mbox{\small loc}}}\left( t \right) = \widehat{\alpha_0^{-1}}(t) * \widehat{I}\left( t \right) + \frac{1}{{\tilde C\left( t \right)}}\int\limits_{ - \infty }^t {I\left( \tau  \right)d\tau }.
\end{equation}
with $\tilde C(t) = C(t)\Delta$. By using  ${1}/(1 + x)  \approx 1 - x$ for $x\ll1$, inserting Eq.~(\ref{capacitors}) in (\ref{E_time}), and transforming to the frequency domain we get,
\begin{equation} \label{freq_domain}
\begin{array}{l}
{E^{\mbox{\small loc}}}\left( \omega  \right) = \,\left[ {{\alpha_0^{-1}}\left( \omega  \right) + \frac{1}{{ j\omega {\tilde{C}_0}}}} \right]I\left( \omega  \right)\\ \\
\! - \frac{{m}}{{2}}\left[ {\frac{{{e^{-j\varphi }}}}{{j\left( {\omega  - \Omega } \right){\tilde{C}_0}}}I\left( {\omega  - \Omega } \right) + \frac{{e^{  j\varphi }}}{{j\left( {\omega  + \Omega } \right){\tilde{C}_0}}}} {I\left( {\omega  + \Omega } \right)} \right].
\end{array}
\end{equation}

Consider now a particular nominal frequency $\omega_0$, which may be, for instance, the excitation frequency of a monochromatic impinging wave, or the excitation frequency of a localized source within the bulk. Then, in light of the harmonic time-modulation the local field  has the following frequency dependence,
\begin{equation} \label{mult_freq_incident_wave}
E^{\mbox{\small loc}}\left( \omega  \right) = \sum\limits_{n =  - \infty }^\infty  E_n^{\mbox{\small loc}}\delta \left( {\omega  - {\omega_n} } \right),
\end{equation}
with $\omega_n = \omega_0 + n\Omega$, and $\delta \left( \cdot \right)$ denotes Dirac's delta.
Then, for the current we have,
\begin{equation} \label{current_shape}
I\left( \omega  \right) = \sum\limits_{n =  - \infty }^\infty  {{I_n}\delta \left( {\omega  - {\omega_n}} \right).}
\end{equation}
Here, ${E_n^{\mbox{\small loc}}}$ and $I_n$ denote the $n$'th harmonic amplitude of the local field and the induced current, respectively. By plugging Eq.~(\ref{mult_freq_incident_wave}) and Eq.~(\ref{current_shape}) into  Eq.~(\ref{freq_domain}), and by balancing the coefficients of equal harmonics we find,
%
%
\begin{equation} \label{E_inc_to_I_n}
\begin{array}{l}
E_n^{\mbox{\small loc}} = \left[ {{\alpha_0^{-1}}\left( {{\omega_n}} \right) + \frac{1}{{j {{\omega_n} } {\tilde{C}_0}}}} \right]{I_n}  \\ \\ -
\frac{m}{2}\left[\frac{e^{-j\varphi}}{j\omega_{n-1}\tilde{C}_0}I_{n-1} + \frac{e^{j\varphi}}{j\omega_{n+1}\tilde{C}_0}I_{n+1} \right].\\
\end{array}
\end{equation}

\subsection{The time-modulated lattice}
Our goal is to analyze a spatiotemporally modulated wire media. While the temporal modulation is introduced by the modulation frequency parameter, $\Omega$, the space modulation is introduced by setting the phase $\varphi$ in Eq.~(\ref{capacitors}) for each of the wires. To that end, we introduce the vector  $\vec \zeta  = \hat x\zeta \cos \xi  + \hat y\zeta \sin \xi $ in the $xy$ plane, that represents the direction and magnitude  at which the capacitors phase $\varphi$ is accumulated. Thus,  in order to achieve the effect of synthetic motion, we choose the phase of the capacitance of the wire with indexes $(m,l)$ to be ${\varphi _{m,l}} = \vec \zeta  \cdot {\vec R}_{m,l} = ma\zeta \cos \xi  + bl\zeta \sin \xi$. By controlling  the phase modulation direction, spatial dispersion is obtained as shown below.
Spatial dispersion has been demonstrated by spatiotemporally modulated metasurface has been shown in \cite{Mazor2019, Mazor2020}. Substantial spatial dispersion may be achieved also in different means  e.g. in \cite{Mode Profile Shaping in Wire Media: Towards An
Experimental Verification} where the authors utilized dielectric rods with different diameter instead of a constant loaded wire.

Due to Floquet-Bloch theorem, the currents  take the form,
\begin{equation}\label{I_m_l_with_perturbation}
{I_{m,l}} = \sum\limits_{n =  - \infty }^\infty  {{A_n}{e^{ - j{{\vec q}_n} \cdot {\vec R}_{m,l}}}\delta \left( {\omega  - {\omega_n}  } \right)}
\end{equation}
where $n$ represents the temporal harmonic number, and the corresponding wave vector is given by (see appendix \ref{sec:Dispersion of the harmonics} for derivation),
\begin{equation} \label{q_n}
{{\vec q}_n} = {{\vec q}_0} + n\vec \zeta
\end{equation}
with ${{\vec q}_0} = \hat x{q_{0,x}} + \hat y{q_{0,y}} = \hat x{q_0}\cos {\theta _0} + \hat y{q_0}\sin {\theta_0}$. Note that ${{\vec q}_0}$ represents the corresponding wave vector at the nominal frequency $\omega_0$. We can now use the expression for the local field, and derive the dispersion relation and  the equations for the eigenmodes. We follow the approach  applied for stationary medium in Sec.~(\ref{Sec_Stationary}), and perform the necessary modifications due to the spatiotemporal modulation of the medium.
Assuming an infinite lattice, with currents given by Eq.~(\ref{I_m_l_with_perturbation}),  the local field on the $(m,l)=(0,0)$ wire reads,
\begin{equation} \label{E_local_media_time_variant}
\begin{array}{l}
{E^{\mbox{\small loc}}} = \sum\limits_{n =  - \infty }^\infty  {E_n^{\mbox{\small loc}}} \delta \left( {\omega  - {\omega_n}} \right)\\
\, =  - \sum\limits_{n =  - \infty }^\infty  {\sum\limits_{m,l \ne 0,0} {
{A_n}G_{nml}
 {e^{ - j{{\vec q}_n} \cdot {\vec R}_{m,l}}}
 \delta \left( {\omega  - {\omega_n}  } \right)} }
\end{array}
\end{equation}
where $G_{nml} = ({{\eta {k_n}}}/{4})H_0^{\left( 2 \right)}\left( {{k_n}{R_{ml}}} \right)$ with $R_{ml}=\sqrt{(ma)^2+(lb)^2}$ and $k_n=\omega_n/c$, denoting the  two dimensional Green's function at the n'th harmonic for a  source located at $\vec R_{ml}$ and observer at $\vec R_{00}$.
By applying Eq.~(\ref{E_local_media_time_variant})  with Eq.~(\ref{freq_domain}), and balancing between equal harmonics at the two sides of the resulting in equation, we find  a recursive relation between the different temporal harmonics of the induced current on the $(m,l)=(0,0)$ wire,
\begin{equation}\label{recurence}
a_nA_{n-1} + b_n A_n + c_n A_{n+1} = 0,
\end{equation}
%
where,
\begin{subequations}\label{anbncn}
\begin{align} \label{a_n_exact}
&{a_n} = \frac{m}{2}\frac{1}{{ {\omega_{n - 1}\tilde{C}_0}}},\\\nonumber\\\label{b_n_exact}
&\begin{array}{l}
{b_n} = \frac{\eta k_n}{2}\left[\frac{1}{\pi}\ln\frac{b}{2\pi r_0} + \frac{1}{b\beta_{n,x}^{(0)}}\frac{\sin \beta_{n,x}^{(0)}a}{\cos \beta_{n,x}^{(0)} - \cos q_{n,x}a}  + \right. \\
\left.  \sum_{l\neq 0}{ \left(\frac{1}{b\beta_{n,x}^{(l)}}\frac{\sin \beta_{n,x}^{(l)}a}{\cos \beta_{n,x}^{(l)} - \cos q_{n,x}a} - \frac{1}{2\pi|l|} \right)}-\frac{1}{\omega_n\tilde{C}_0}  \right],
\end{array}\\\nonumber\\\label{c_n_exact}
&c_n=\frac{m}{2}\frac{1}{{ {\omega_{n + 1}\tilde{C}_0}}},
\end{align}
\end{subequations}
with $\beta_{n,x}^{(l)}=-j\sqrt{\left( q_{n,y}+\frac{2\pi l}{b} \right)^2-k_n^2},\, \mbox {Re} \left\{ \sqrt{\cdot}\right\}\!>\!0$.
For a dense grid, as in Sec.~\ref{Sec_Stationary},  we  approximate $b_n$ in Eq.~(\ref{b_n_exact}),
%
%
%
\begin{equation}\label{b_n_approx}
b_n = \omega_n\tilde{L} - \frac{1}{\omega_n\tilde{C}_0}-\frac{\eta k_n}{ab(k_n^2-q_n^2)}
\end{equation}
with,
\begin{equation} \label{q_n_sqr}
q_n^2 = {\left( {{{\vec q}_0} + n\vec \zeta } \right)^2} = q_0^2 + 2n\zeta {q_0}\cos \gamma  + {n^2}{\zeta ^2}.
\end{equation}
Here, $(\vec{x})^2=\vec{x}\cdot\vec{x}$,  and  $\gamma=\theta_0-\xi$. In Eq.~(\ref{b_n_approx}),  $\tilde{C}_0 =C_0\Delta $ as before and
\begin{equation}\label{L_tilda_media}
\tilde{L} = \tilde{L}_w + \tilde{L}_0 \mbox{ with } \tilde{L}_0=\frac{L_0}{\Delta}
\end{equation}
%
where $L_0$ is an inductance that may be connected in series with the loading capacitor $C(t)$. Effectively it simply increases the intrinsic wire inductance. Note that as  in the case of stationary wire medium with dense grid, also here the approximation for $b_n$ in Eq.~(\ref{b_n_approx}) depends only on $q_n^2$, whereas the exact  expression for $b_n$ in Eq.~(\ref{b_n_exact}) depends also on $q_{n,x}$.

\begin{figure}[ptb]
\centering
\vspace{-0.0cm}  \includegraphics[width=8cm]{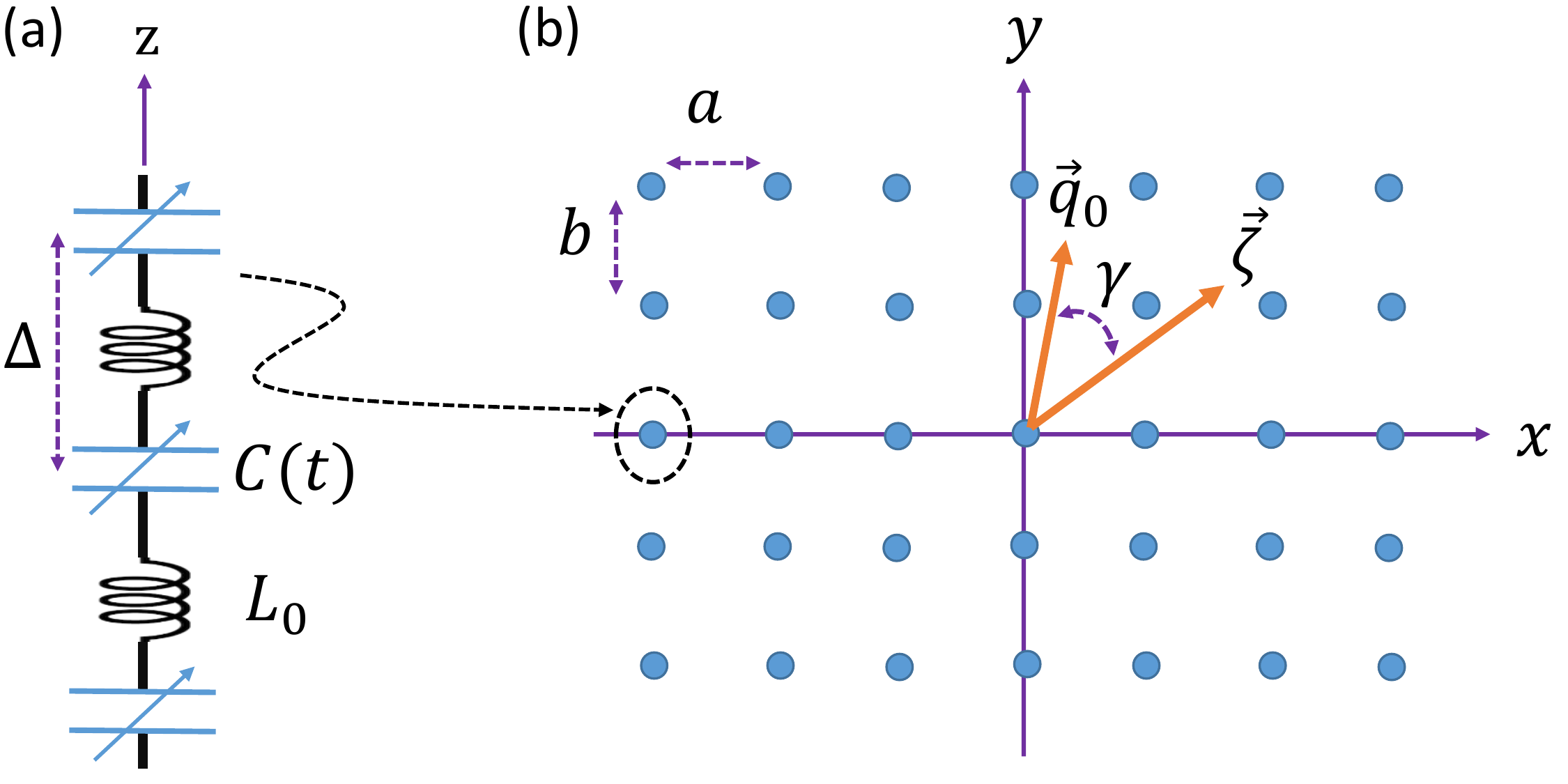} \vspace{-0.0cm}%
\caption{(a) The time-modulated loaded wire. $L_0$ increases the self inductance of the wire, and $C(t)$ introduces the time-modulated capacitance. These lumped elements are periodically loaded on the wire with periodicity $\Delta$, thus creating the effective wire time-modulated susceptibility. (b) The wire medium. Each wire is modulated with the same temporal frequency $\Omega$, but with a different phase that is determined by the modulation wave vector $\vec{\zeta}$. The propagation takes place with a fundamental wave vector $\vec{q}_0$. }
\label{wire_media_L_and_C}%
\end{figure}


%
%
As opposed to the stationary loaded wire medium case where the effective longitudinal permitivity may be analytically expressed using Eq.~(\ref{eps_LTI}) with Eq.~(\ref{Plasma_Freq_LTI}), for the spatiotemporally modulated case, the interaction between the different temporal frequency harmonics should be  included. To that end, Eq.~(\ref{recurence}) is used. The latter can be represented using a tridiagonal matrix of infinite rank. We assume that  for large enough $N$, the currents $A_{-\left( N+1 \right)},A_{N+1}$ are negligible. In this case the infinite matrix  can be  approximated by a finite square matrix %
\begin{equation} \label{det_A_demand}
{\underline{\underline A} ^N} = \left[ {\begin{array}{*{20}{c}}
 \ddots & \ddots &0&0&{}&{}&{}\\
 \ddots &{{b_{ - 2}}}&{{c_{ - 2}}}&0&0&{}&{}\\
0&{{a_{ - 1}}}&{{b_{ - 1}}}&{{c_{ - 1}}}&0&0&{}\\
0&0&{{a_0}}&{{b_0}}&{{c_0}}&0&0\\
{}&0&0&{{a_1}}&{{b_1}}&{{c_1}}&0\\
{}&{}&0&0&{{a_2}}&{{b_2}}& \ddots \\
{}&{}&{}&0&0& \ddots & \ddots
\end{array}} \right].
\end{equation}
For nontrivial wave solutions we look for the dispersion relation $\vec{q}_0(\omega)$ that nullifies the determinant,
\begin{equation}\label{detAN}
 |\underline{\underline{A}}^N|=0.
\end{equation}
In the following we explore the wave dynamics in the bulk as obtained from this dispersion relation  for a particularly interesting case near the plasma frequency of the corresponding stationary wire medium.

\section {Near the plasma frequency of the  stationary medium}\label{Sec Near Plasma}

In the following we explore the wave dynamics of the spatiotemporally modulated medium with parameters that are near the plasma frequency of the corresponding stationary medium. Namely, with $\omega = \omega^0 + \delta \omega$, where $\omega^0$ is the frequency that nullifies $\epsilon_z(\omega)$ in Eq.~(\ref{eps_LTI}). Thus
\begin{equation} \label{k_0_sqr}
\omega^0 = \sqrt{\frac{1}{\tilde{L}}\left(\frac{1}{\tilde{C}_0}+\frac{1}{\tilde{C}_b}\right)} = \sqrt{\frac{1}{{\tilde{L}\tilde{C}_0}}}\sqrt{1+\psi}
\end{equation}
where $C_b$ is defined after Eq.~(\ref{Plasma_Freq_LTI}), and
%
%
%

%
\begin{equation} \label{psi_definition}
 \psi  = \frac{\tilde{C}_b}{\tilde{C}_0}.
\end{equation}
In the following we will also use $k^0=\omega^0/c$.

\subsection{Analytic approximation for the lower order solutions with weak modulation}
Assuming weak modulation, $m\ll1$, and as a result the dominant harmonics beside the fundamental one will be $n=\pm1$. Note that this assumption excludes the possibility that higher order harmonic will be strongly excited in this case due to the presence of some resonance mechanism at some other frequency. Under this assumption the infinite rank matrix $\underline{\underline{A}}$ will be replaced by a tridiagonal $3\times3$ matrix $\underline{\underline{A}}^1$ ($N=1$ in Eq.~(\ref{detAN})). The dispersion relation in this case is found by nullifying its determinant, i.e.,
%


%
\begin{equation} \label{approximate_determinant_equation}
\frac{{{a_0}{c_{ - 1}}}}{{{b_{ - 1}}}} + \frac{{{a_1}{c_0}}}{{{b_1}}} = {b_0}.
\end{equation}
By plugging the expressions for $a_n$ and $c_n$ that are given in Eq.~(\ref{anbncn}), with the dense grid approximation Eq.~(\ref{b_n_approx}), together with Eq.~(\ref{psi_definition}), and with $q_1,q_{-1}$ given in Eq.~(\ref{q_n}), the non-trivial solution requirement in Eq.~(\ref{approximate_determinant_equation}) turns into  a sixth order polynomial equation for $q_0$,
%
%
\begin{subequations}\label{q_0_equation}
\begin{equation}\label{q_0_equation_with_psi}
\frac{m^2\psi^2}{4} \left[\frac{1}{X_{-1}(\omega,q_0)} + \frac{1}{X_1(\omega,q_0)} \right] = X_0(\omega,q_0)
\end{equation}
with
\begin{equation}
X_n(\omega,q_0) = \frac{k_n^2}{\left(k^0\right)^2} \left(1+\psi \right) - \psi - \frac{k_n^2}{k_n^2-q_n^2}.
\end{equation}
\end{subequations}
%
Let us the denote the $n$'th root of Eq.~(\ref{q_0_equation_with_psi}) by $q_0^{\#n}$ where $n=1..6$. Two of the six solutions of this polynomial equation will represent \emph{a perturbation over the two solutions of the corresponding stationary wire medium}. Since we focus here on the dispersion near the plasma frequency of the corresponding stationary medium, i.e., in the vicinity of $\omega^0$, it is reasonable to assume that two of the solutions satisfy ${{q_0^{\# 1}}},\,{{q_0^{\# 2}}} \ll k^0$. In that case, in the left-hand-side of Eq.~(\ref{q_0_equation_with_psi}) we will use $\omega=\omega^0$, and $q_0=0$, and thus $X_1,X_{-1}\sim 1$.  Since $m\ll1$,
the plasma frequency of the spatiotemporally modulated medium, namely, the frequency at which $q_0^{\#1,2}$ experience the transition from being purely imaginary to purely real shifts by $\delta\hat\omega$ with respect to $\omega^0$ of the stationary medium where,
%
\begin{equation}
\frac{{\delta \hat \omega }}{{{\omega ^0}}} \approx \left[\frac{1}{X_1(\omega^0,0)} + \frac{1}{X_{-1}(\omega^0,0)} \right]\,\frac{{{m^2}{\psi ^2}}}{{8\left( {1 + \psi } \right)}}.
\end{equation}
%
Then, the first two solutions $q_0^{\#1,2}$ for the fundamental modes, up to the shift in $\delta\hat\omega$ are obtained by solving $X_0(\omega,q_0)=0$ and are given by,
%
\begin{equation}
\begin{array}{l} \label{q_0_1}
q_0^{\# 1,2} = \pm{k^0}\left( {1 + \frac{{\delta \omega}}{{{\omega ^0}}} - \frac{{\delta \hat \omega }}{{{\omega ^0}}}} \right) \times \\ \\
\,\,\,\,\,\,\,\,\,\,\,\,\,\,\,\,\,\,\,\,\,\sqrt {1 - \frac{1}{{{{\left( {1 + \frac{{\delta \omega }}{{{\omega ^0}}} - \frac{{\delta \hat \omega }}{{{\omega ^0}}}} \right)}^2}\left( {1 + \psi } \right) - \psi }}}.
\end{array}
\end{equation}
We can see that under the dense grid approximation, solutions ${\# 1,\# 2}$ are isotropic, independent of the relative angle $\gamma$ between spatiotemporal modulation axis and the direction of  the wave propagation.

\begin{figure}[ptb]
\centering
\vspace{-0.0cm} \includegraphics[width=8cm]{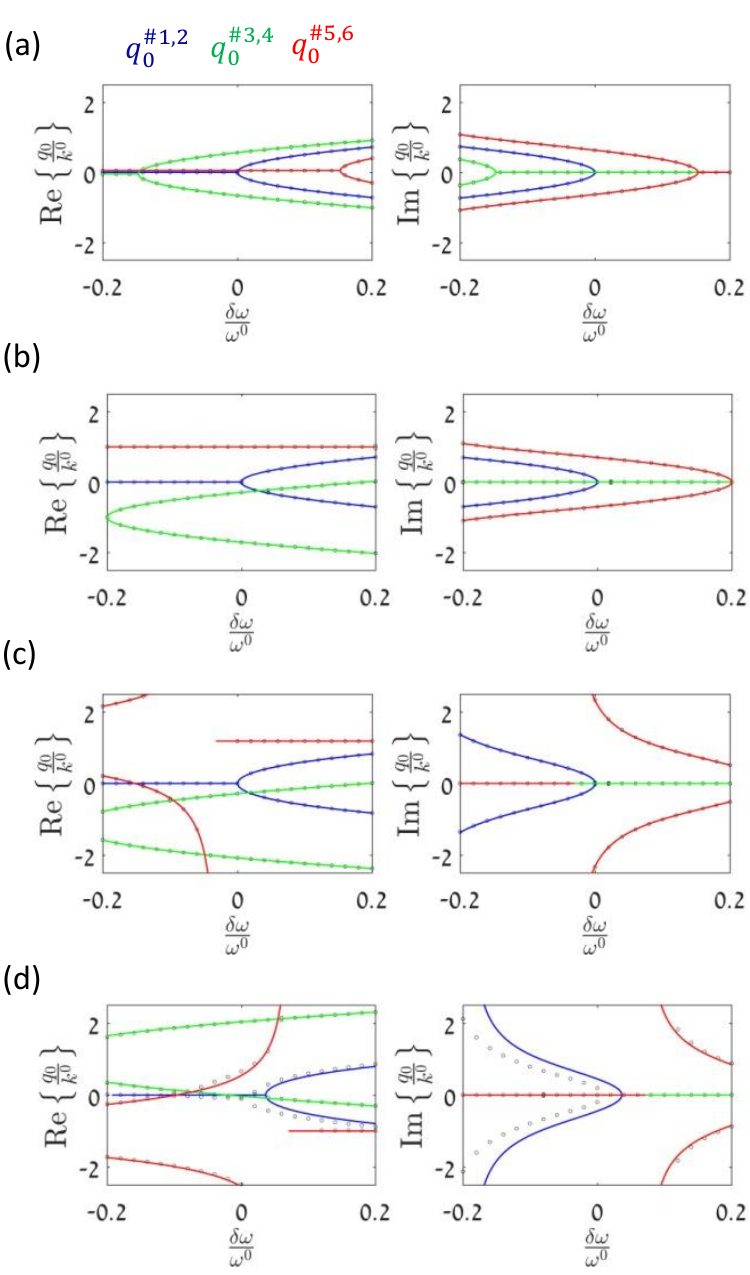} \vspace{-0.0cm}%
\caption{Dispersion relation as obtained by solutions \#1-\#6 in Eqs.~(\ref{q_0_1})-(\ref{q_n_46}). The fundamental wavenumber $q_0$ of each one of the solutions is shown as a function of the frequency $\delta\omega/\omega_0$ for the four parameter cases that are tabulated in Table 1. (a) Case I, (b) Case II, (c) Case III, and (d) Case IV.
}
\label{Fig3}
\end{figure}

To find the other four roots of Eq.~(\ref{q_0_equation_with_psi}) we shall assume that these solutions are not small compare to $k^0$ near $\omega^0$.  As a result, the right-hand-side of Eq.~(\ref{q_0_equation_with_psi}) $X_0(\omega^0,q_0)\sim 1$. On the other hand, in light of the small factor $m^2$ in the left-hand-side, the only way to balance the two sides is if either $X_1$ or $X_{-1}$ are nearly vanishing.   Therefore   the remaining four roots are approximated by solving $X_{1}(\omega,q_0)=0$ and $X_{-1}(\omega,q_0)=0$. Each one of the two equations yields two solutions,
\begin{subequations}\label{q_n_46}
\begin{alignat}{3}
&\begin{array}{l}\label{q_34}
q_0^{\# 3,4} =  - \zeta \cos \gamma  \pm \\ \\
\,\,\,\,\,\,\,\,\,\,\,\,\,\,{k_1}\sqrt { - {{\left( {\frac{\zeta }{{{k_1}}}} \right)}^2}{{\sin }^2}\gamma  + 1 - \frac{1}{{\frac{{k_1^2}}{{{{\left( {{k^0}} \right)}^2}}}\left( {1 + \psi } \right) - \psi }}}
\end{array} \\ \nonumber \\
&\begin{array}{l}\label{q_56}
q_0^{\# 5,6} = \zeta \cos \gamma  \pm \\ \\
\,\,\,\,\,\,\,\,\,\,\,\,\,{k_{ - 1}}\sqrt { - {{\left( {\frac{\zeta }{{{k_{-1}}}}} \right)}^2}{{\sin }^2}\gamma  + 1 - \frac{1}{{\frac{{k_{ - 1}^2}}{{{{\left( {{k^0}} \right)}^2}}}\left( {1 + \psi } \right) - \psi }}}
\end{array}
\end{alignat}
\end{subequations}
If imaginary, the square roots in  Eqs. (\ref{q_0_1}) and (\ref{q_n_46}) should be chosen to guarantee physical solutions, namely, the wave must decay at infinity.
In Fig.~\ref{Fig3} we show the dispersion of the six solutions as obtained by the analytical approximated relations in Eqs.~ (\ref{q_0_1}) and (\ref{q_n_46}), the dispersion is shown for the following structure parameters $a=0.07\lambda^0, b=0.1\lambda^0, r_0=0.0001\lambda^0$, and $\theta_0=0$. As opposed to the structural parameters, the \emph{modulation} parameters differ between four typical cases that are summarized in the table below.
\begin{table}
\begin{tabular}{|l|c|c|c|c|c|c|c|}
  \hline
           & $m=\delta C/C_0$ & $\Omega/\omega^0$ & $\zeta/k^0$ & $\tilde{L}_0/\tilde{L}_w$  & $\xi$ & $\psi$ \\ \hline
  Case I   & 0.1  & 0.15 & 0.1 & 4.3 & $-\pi/3$  & 0.27 \\\hline
  Case II  & 0.35 & 0.2  & 1.0 & 4.0  & 0 & 0.2 \\\hline
  Case III & 0.06 & 0.25 & 1.2 & 7.6 & $\pi/20$ &  1.07 \\\hline
  Case IV & 0.25 & 0.3 & 1.0 & 9.3 & $-\pi$ &  1.47 \\
  \hline
\end{tabular}
\caption{Four parameter cases used in the numerical calculations in this section.}\label{Table1}
\end{table}
The complex dispersion diagrams, real and imaginary $q_0$, are shown as a function of $\delta\omega$ which is the detuning frequency from $\omega^0$ - the plasma frequency of the stationary lattice. In all four cases the lines are color-coded, blue for $q_0^{\#1,2}$ - the perturbation on the solution of the stationary medium, whereas green and red lines are used for the new solutions due to the spatiotemporal modulation, $q_0^{\#3,4}$ and $q_0^{\#5,6}$, respectively. The colored lines in Fig.~\ref{Fig3} are obtained by using the approximated solutions that are given in Eqs.~(\ref{q_0_1}) and (\ref{q_34},\ref{q_56}), to show the analytic approximation validity, as a comparison in the figures we also draw by black circles the numerical solution of Eq.~(\ref{q_0_equation_with_psi}).

The good agreement between the analytical approximations, and the numerical calculations, allow us to use our analytical results for synthesis. For example, an interesting characteristic for the roots of  $q_0^{\# 3},q_0^{\# 4},q_0^{\# 5}$ and $q_0^{\# 6}$, is that we can create a different response for different angle $\gamma$ and in particular reach effective Fresnel drag dispersion accompanied with strong anisotropy for waves propagating in the wire medium bulk by using \emph{slow} spatiotemporal modulation.

\subsection{Effective Fresnel drag with weak and slow spatiotemporal modulation}
The spatiotemporal modulation enables to reach a dispersion that shares similarities with the dispersion of waves in a moving medium. However, typically this effect at the fundamental harmonic  frequency is rather weak and requires  time modulation much faster then the guided signal frequency \cite{Fresnel drag in space time modulated metamaterials}. This is not the case however when we  spatiotemporally modulating the loaded wire medium. In this section we show that it is the \emph{interplay} between the \emph{spatiotemporal modulation} and the \emph{cutoff of the fundamental harmonic} below $\omega^0$ - the plasma frequency of the corresponding stationary medium - that enables to obtain effectively Fresnel drag with no actual motion and using slow and weak modulation.
To see this effect, we use Eq.~(\ref{q_0_1}), and Eqs.~(\ref{q_34},\ref{q_56}) in order to extract the relation between $\delta\omega$ and the transverse wave number components $q_{0x} = q_0\cos\theta_0$ and $q_{0y} = q_0\sin\theta_0$ for a given set of medium and modulation parameters.
By solving Eq.~(\ref{q_0_1}) for $\delta\omega$ we obtain,
\begin{equation}
{\frac{{\delta \omega }}{{{\omega ^0}}}^{\#1,2}} = \frac{{\delta \hat \omega }}{{{\omega ^0}}} + \frac{{\frac{{q_0^2}}{{2{{\left( {{k^0}} \right)}^2}}}}}{{\left[ {1 + \psi \left( {1 - \frac{{q_0^2}}{{{{\left( {{k^0}} \right)}^2}}}} \right)\,} \right]}}
\end{equation}
whereas by solving  Eqs.~(\ref{q_34}) and (\ref{q_56}) for $\delta\omega$ we get,
\begin{equation}\label{omegapm}
{\frac{{\delta \omega }}{{{\omega ^0}}}^{\pm1}} = \frac{{1 + \left( {1 - \frac{\psi }{{1 + \psi }}} \right)\frac{{q_{\pm1}^2}}{{{{\left( {{k^0}} \right)}^2}}} - {{\left( {1 \pm \frac{\Omega }{{{\omega ^0}}}} \right)}^2}}}{{4{{\left( {1 \pm \frac{\Omega }{{{\omega ^0}}}} \right)}^2} - 2\left[ {1 + \frac{{q_{\pm1}^2}}{{{{\left( {{k^0}} \right)}^2}}}} \right]}}
\end{equation}
where the $+1$ [$-1$] corresponds to solutions \#3 and \#4 [\#5 and \#6] in Eqs.~(\ref{q_n_46}), and $q_{\pm1}$ is given using Eq.~(\ref{q_n_sqr}). In the latter we note that $q_0^2=q_x^2+q_y^2$, $\gamma = \theta_0-\xi$ where $\xi$ is the direction of the modulation and  $\tan\theta_0 = q_y/q_x$.
\begin{figure}[ptb]
\centering
\vspace{-0.0cm} \includegraphics[width=7cm]{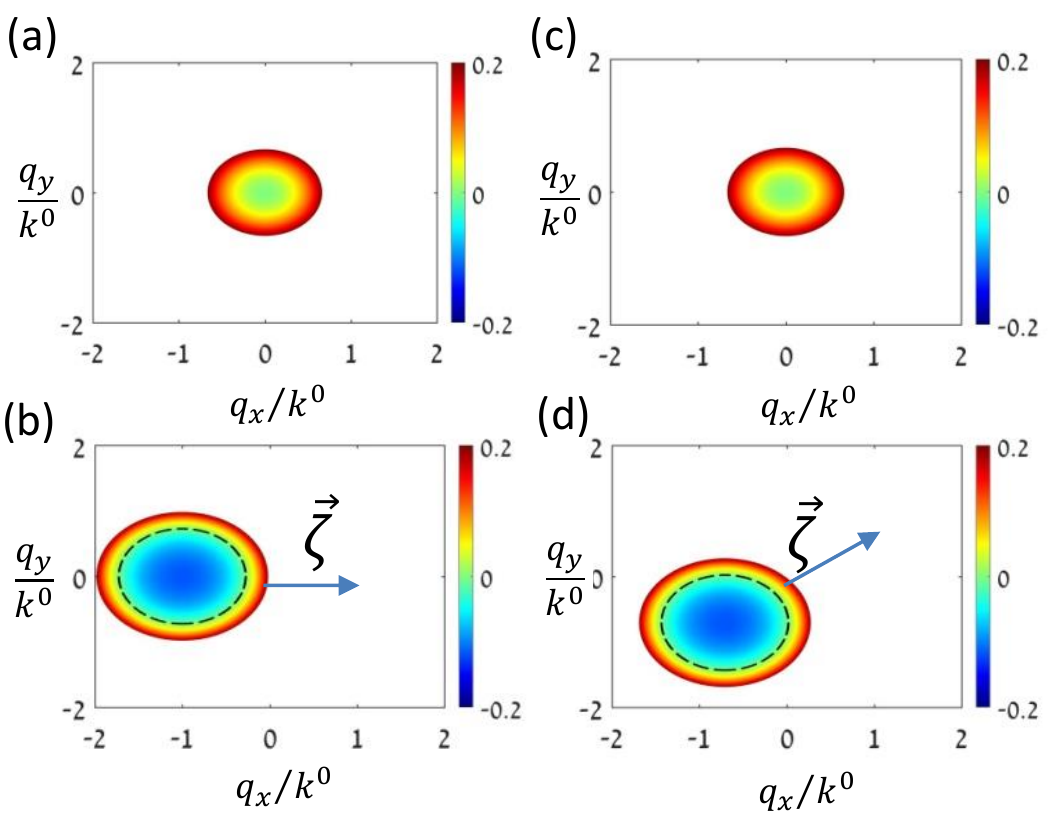} \vspace{-0.0cm}%
\caption{Dispersion diagram of the fundamental solutions \#1 and \#2 and high order solutions \#3-\#4. (a) With $\xi=0$ the fundamental harmonic dispersion is centered at the origin of the ($q_x,q_y$) plane, however, its cutoff frequency is $\delta\omega/\omega_0\approx 0$. (b) As opposed to that, the high order solutions are propagating already at negative frequencies. These solutions are essentially nonreciprocal in light of their asymmetric dispersion. (c) and (d) are as (a) and (b) but with $\xi=\pi/4$.
}
\label{Fig4}
\end{figure}
These dispersion relations are shown in Fig.~\ref{Fig4} below. In (a) with $\xi=0$, a color map that shows $\delta\omega/\omega_0$ as a function of the transverse wave vector $(q_x,q_y)$ is presented, showing that the fundamental solutions are propagating only at positive frequencies (to be more precise $\frac{{\delta \omega}}{{{\omega ^0}}}^{\#1,2} = \frac{{\delta \hat \omega }}{{{\omega ^0}}}\approx 0$). However, the $\pm1$ harmonics (see Eq.~(\ref{omegapm})), with strongly \emph{asymmetric} dispersion, are propagating at \emph{negative} frequencies, as shown in Fig.~\ref{Fig4}(b). As a result, in the spatiotemporally modulated wire medium effective motion may be exhibited at the first propagating harmonic and with slow modulation. This behaviour may be controlled by adjustments of the direction of modulation, as shown in Fig.~\ref{Fig4}(c) for the fundamental harmonic, and in (d) for the $\pm1$ harmonic, this time with spatiotemporal modulation along the $\xi=45\deg$ direction.

\section{Homogenization}\label{Sec Homo}
\subsection{Effective modal permittivity}
\begin{figure}[ptb]
\centering
\vspace{-0.0cm}
\includegraphics[width=\columnwidth]{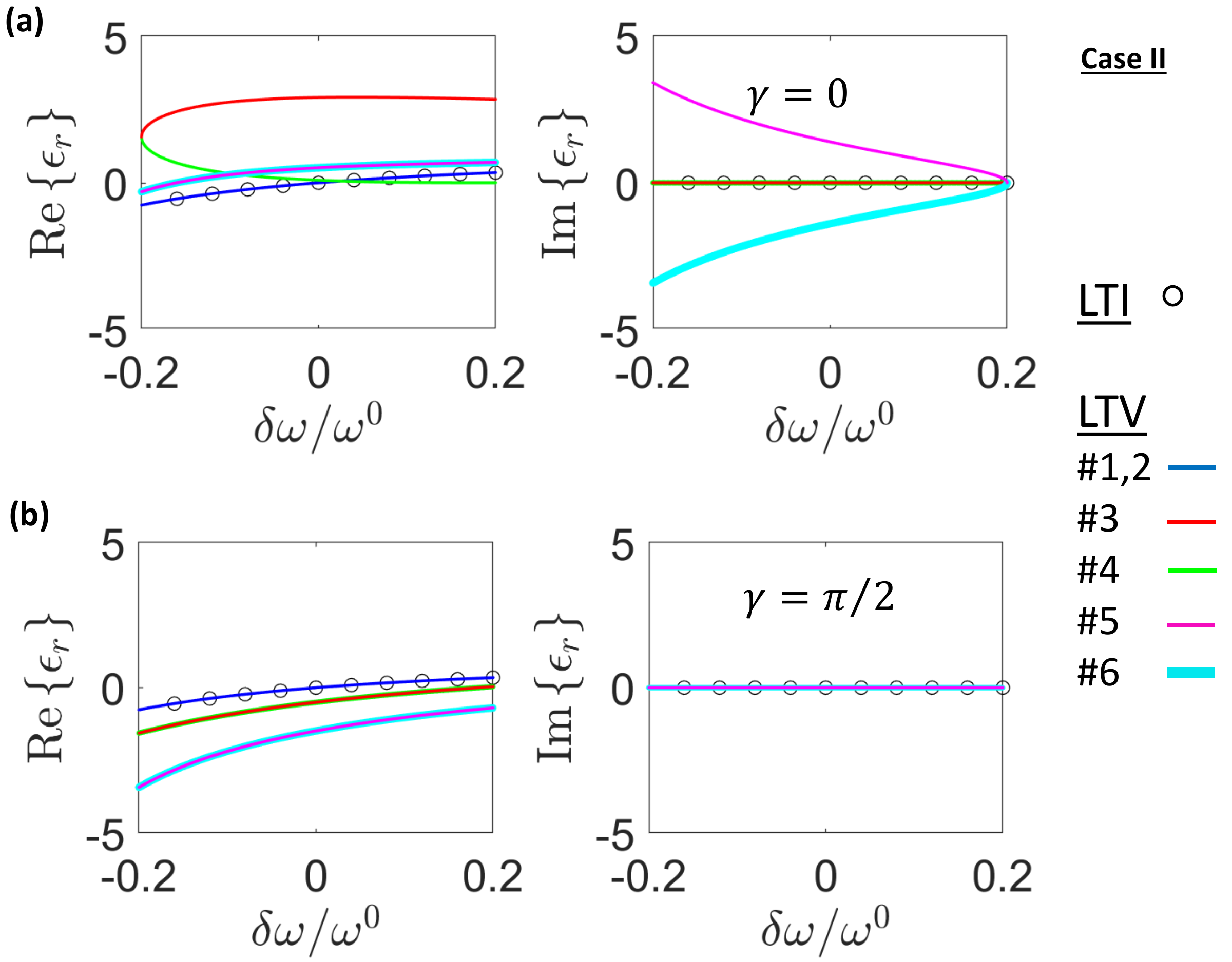}
\vspace{-0.0cm}%
\caption{Effective permittivity of spatiotemporally modulated wire medium. Parameters of case II in Table \ref{Table1}. (a) effective permittivity of the modal solutions when the propagation is collinear with the modulation. Circles: no modulation - stationary wire media. Cont. lines: blue-fundamental modal solutions \#1,2 are nearly identical to these of the stationary medium, and are reciprocal. The other higher order solutions \#3,4 and \#5,6  exhibit deviation with respect to the stationary medium solutions and substantial nonreciprocally. (b) as (a) but for propagation normal to the direction of modulation. The solutions are reciprocal in this case, and thus the strong anisotropy due to the space-time modulation clearly emerges by comparing the two figures (a) and (b).}
\label{Fig_EffectivePermittivity_wire_case2}%
\end{figure}

\begin{figure}[ptb]
\centering
\vspace{-0.0cm}
\includegraphics[width=\columnwidth]{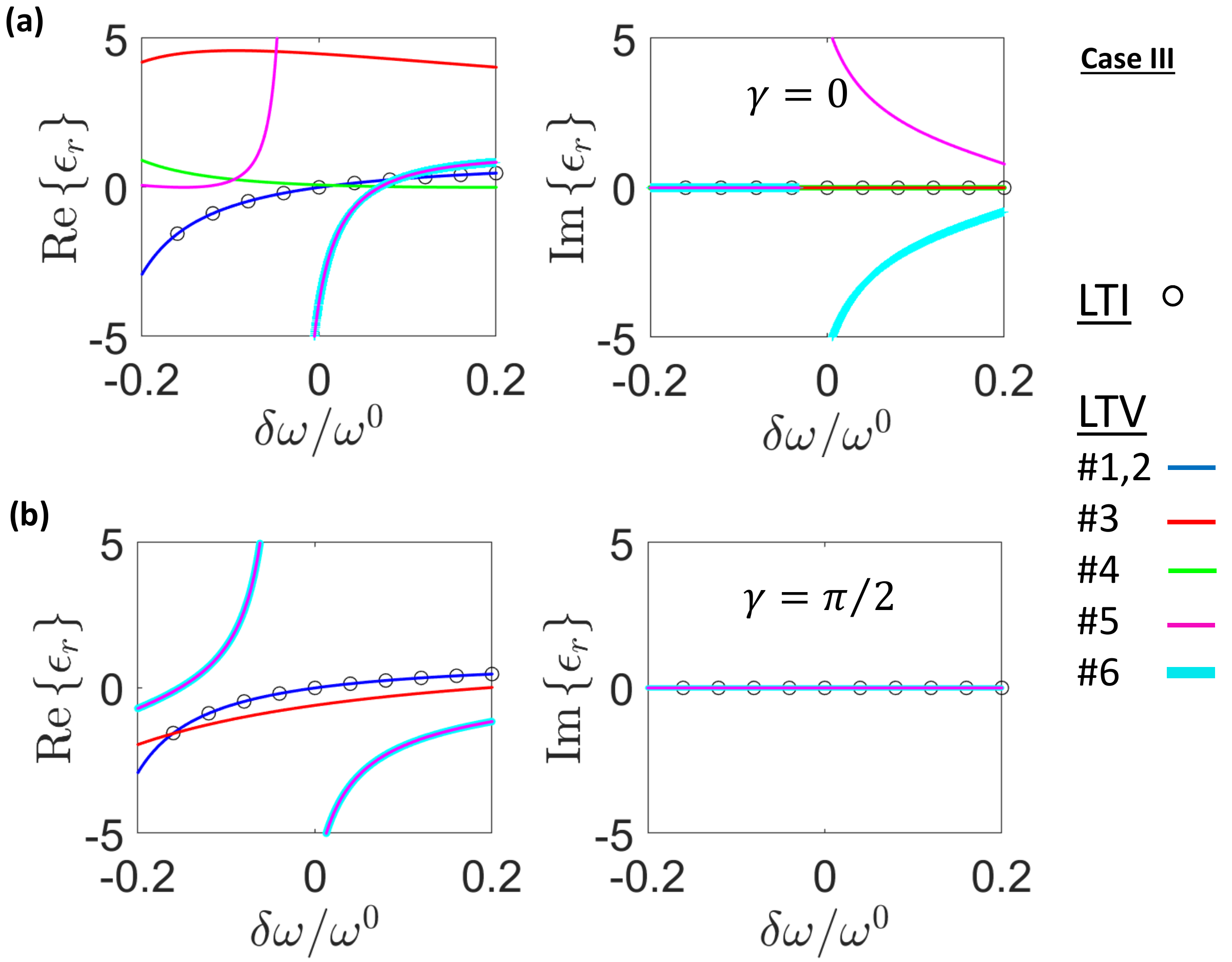}
\vspace{-0.0cm}%
\caption{Effective permittivity of the modal solutions for case III in Table~\ref{Table1}. As in Fig.~\ref{Fig_EffectivePermittivity_wire_case2} except that the fact that $\psi\sim1$ implies much stronger effect by the modulation as evident by comparing with Fig.~\ref{Fig_EffectivePermittivity_wire_case2}.}
\label{Fig_EffectivePermittivity_wire_case3}%
\end{figure}

Using the wavenumber dispersion of the modal solutions given in Eqs.~(\ref{q_0_1})-(\ref{q_n_46}) it is possible to define effective permittivity for each one of the modal solutions found above.   To that end we first write free space wavenumber as
\begin{equation}
{{k^2}}  = {\left( {{k^0}} \right)^2}{\left( {1 + \frac{{\delta \omega }}{{{\omega ^0}}}} \right)^2}.
\end{equation}
Then, the homogenized effective permittivity of the modal solution \#$i$ is given by
\begin{equation}
\varepsilon_{r:i} = {\left( {q_0^{\# i}} \right)^2}/{k^2}.
\end{equation}
Note that this effective relative permittivity connects the free space wave number with the \emph{fundamental} guided mode wavenumber of each one of the modal solutions. It is important to stress that each one of these modes, with $q_0^{\# i}$, consists of infinite space-time harmonics as dictated by Eq.~(\ref{I_m_l_with_perturbation}). Once $\varepsilon_{r:i}$ is known, $q_0^{\#i}$ can be calculated. Then, using Eq.~(\ref{q_n}), the wavenumber of each one of the infinite harmonics that consists the $i$ modal solution is immediately found.
Specifically, for the first six fundamental modal solutions in the spatioteporally modulated wire medium we have,
\allowdisplaybreaks
\begin{widetext}
\begin{subequations}
\begin{align}
&{\varepsilon _{r:1}} = {\varepsilon _{r:2}}= {\left( {\frac{{1 + \frac{{\delta \omega }}{{{\omega ^0}}} - \frac{{\delta \hat \omega }}{{{\omega ^0}}}}}{{1 + \frac{{\delta \omega }}{{{\omega ^0}}}}}} \right)^2}\left[ {1 - \frac{1}{{{{\left( {1 + \frac{{\delta \omega }}{{{\omega ^0}}} - \frac{{\delta \hat \omega }}{{{\omega ^0}}}} \right)}^2}\left( {1 + \psi } \right) - \psi }}} \right]\label{eps_LTV_1}\\
&{\varepsilon _{r:3,4}} = \frac{1}{{{{\left( {1 + \frac{{\delta \omega }}{{{\omega ^0}}}} \right)}^2}}}{\left[ { - \frac{\zeta }{{{k^0}}}\cos \gamma  \pm \frac{{{k_1}}}{{{k^0}}}\sqrt { - {{\left( {\frac{\zeta }{{{k_1}}}} \right)}^2}{{\sin }^2}\gamma  + 1 - \frac{1}{{\frac{{k_1^2}}{{{{\left( {{k^0}} \right)}^2}}}\left( {1 + \psi } \right) - \psi }}} } \right]^2}\\
&{\varepsilon _{r:5,6}} = \frac{1}{{{{\left( {1 + \frac{{\delta \omega }}{{{\omega ^0}}}} \right)}^2}}}{\left[ {\frac{\zeta }{{{k^0}}}\cos \gamma  \pm \frac{{{k_{ - 1}}}}{{{k^0}}}\sqrt { - {{\left( {\frac{\zeta }{{{k_{ - 1}}}}} \right)}^2}{{\sin }^2}\gamma  + 1 - \frac{1}{{\frac{{k_{ - 1}^2}}{{{{\left( {{k^0}} \right)}^2}}}\left( {1 + \psi } \right) - \psi }}} } \right]^2}
\end{align}
\end{subequations}
\end{widetext}
As evident by Eq.~(\ref{eps_LTV_1}), clearly the effective permittivities that correspond with the first two modal solutions exhibit a reciprocal behaviour and are independent of the direction of propagation, and moreover they become positive only above $\delta\omega/\omega_0\approx 0$. This is what we expect from the a \emph{stationary} wire medium as given in Eq.~(\ref{eps_LTI}).  This is reasonable since these two solutions in the spatiotemporally modulated wire media are essentially a weak \emph{perturbation} over the solutions of the conventional stationary wire media. Note however that by comparing the terms in Eq.~(\ref{eps_LTI}) and Eqs.~(\ref{eps_LTV_1}), it is clear that the modulation parameters affect to some extent the values of the permittivity, albeit not its general properties. Moreover, while Eq.~(\ref{eps_LTV_1}) exhibits a reciprocal, stationary-medium-like, behaviour, there are additional higher order harmonics (see Eq.~(\ref{q_n})) that break the reciprocity.
As opposed to $\varepsilon_{r:1,2}$, the additional effective permittivities, i.e.,  $\varepsilon_{r:3}$ and on, demonstrate a substantial nonreciprocity already at the fundamental harmonic. Even more important is the fact that these solutions are propagating already with negative frequencies $\delta\omega/\omega_0<0$ in the region where the first two solutions are evanescent. This implies, as we already discussed in the previous section, in the context of Fresnel drag, that a substantial nonreciprocity can be observed in the spatiotemporally modulated wire media already with weak and slow modulation parameters. We would like also to stress  that these modes are propagating bellow the plasma frequency \emph{for any modulation depth} $m$, \emph{even if extremely small}. Therefore, these modes share similarities with the extraordinary Whistler mode in magneto-plasma  that also propagates below the plasma frequency with any magnetic biasing \cite{Ishimaru}.
This behaviour is shown in Fig.~\ref{Fig_EffectivePermittivity_wire_case2} that shows for the parameters that are given case II of Table~\ref{Table1} a plot of the effective permittivity for the first six fundamental modal solutions, as a function of $\delta\omega/\omega^0$ and  for two main situations. In (a,b),  real and imaginary parts, where the propagation is parallel to the modulation axis, $\vec{q}_0||\vec{\zeta}$, and  in (c,d) where $\vec{q}_0\perp\vec{\zeta}$.
The effective permittivity for the stationary (LTI medium) is shown in black circles, while the solutions for the spatiotemporally modulated lattice are shown in the colored lines.
In Fig.~\ref{Fig_EffectivePermittivity_wire_case2} the high anisotropy is evident due to the directional preference dictated by the spatiotemporal modulation vector $\vec{\zeta}$. In addition,  the two fundamental solutions \#1,2 are essentially identical to the two counter-propagating solutions of the corresponding stationary (LTI) medium (shown in black circles), and as such they  propagate only with $\delta\omega>0$. In contrast, the higher order solutions may be highly nonreciprocal, except for the case where the propagation is transverse to the modulation vector (c,d). As an example, see the counter propagating solutions \#3 and \#4 in Fig.~\ref{Fig_EffectivePermittivity_wire_case2}(a) that are associated with different effective permittivities. Moreover, it is seen  that the spatiotemporal modulation yields a magnetized plasma-like extraordinary wave propagation in the sense that propagation becomes alowed with any modulation index below the so-called plasma-frequency $\omega^0$.
As opposed to Fig.~\ref{Fig_EffectivePermittivity_wire_case2} that uses $\psi=0.2\ll1$, in Fig.~\ref{Fig_EffectivePermittivity_wire_case3} we show similar results but with larger capacitive loading, so that $\psi=1.07\sim 1$. Here, the effect of the spatiotemporal modulation becomes stronger, as evident by a comparison between Fig.~\ref{Fig_EffectivePermittivity_wire_case2} and Fig.~\ref{Fig_EffectivePermittivity_wire_case3}. Also here, the fundamental solutions \#1 and \#2 exhibit permittivity that is essentially identical to that of the corresponding unmodulated lattice. However, as opposed to Fig.~\ref{Fig_EffectivePermittivity_wire_case2}, here, the extraordinary modes that are supported below the plasma frequency are much more dispersive, and the non-reciprocity becomes more dominant, as can be viewed by comparing the counter propagating solution pairs \#3,4 and \#5,6. We will show in Sec.~\ref{Sec perturbed cont} that as opposed to the effective permittivities that are shown in Fig.~\ref{Fig_EffectivePermittivity_wire_case3} for parameters case III, the effective permittivities shown in Fig.~\ref{Fig_EffectivePermittivity_wire_case2} for parameters case II can be adequately reproduced in a model in which the spatiotemporal modulation is introduced after the homogenization.

\subsection{Averaged field equations and Poyinting's vector}
For each one of the the modal solutions, with $q_0=q_0^{\#i}$, we write the governing electrodynamic equations for the \emph{averaged} fields.
We begin by calculating the average fields in the unit-cell at the origin. To that end, we use the expression for the currents in Eq.~(\ref{I_m_l_no_perturbation}). For the fundamental harmonic of each one of the modal solutions, the electric field can be written in the same way as in Eq.~(\ref{E_local_media}), but with the modification for the entire unit cell, except on the wire itself,
\begin{equation} \label{E_unit_cell_origin}
\begin{array}{l}
\vec E\left( {\vec r} \right) =  - \frac{{\hat z\eta k}}{4}\sum\limits_{m,l} {H_0^{\left( 2 \right)}\left( {k\left| {{{\vec R}_{ml}} - \vec r} \right|} \right){I_0}{e^{ - j\left( {{q_x}am + {q_y}bl} \right)}}} \\
\,\,\,\,\,\,\,\,\,\,\,\,\, =  - \frac{{\hat z\eta k{I_0}}}{2}\sum\limits_{m =  - \infty }^\infty  {{S_m}\left( {\vec r} \right){e^{ - j{q_x}am}}}
\end{array}
\end{equation}
with,
\begin{equation}\label{EqSm}
\begin{array}{l}
{S_m}\left( {\vec r} \right) = \frac{1}{2}\sum\limits_{l =  - \infty }^\infty  {H_0^{\left( 2 \right)}\left( {k\left| {{{\vec R}_{ml}} - \vec r} \right|} \right){e^{ - j{q_y}bl}}} \\
\,\,\,\,\,\,\,\, =  - \frac{1}{b}\sum\limits_{l =  - \infty }^\infty  {\frac{{{e^{jy\left( {{q_y} + \frac{{2\pi l}}{b}} \right)}}{e^{ - ja\left| {m - \frac{x}{a}} \right|{\beta _{x,l}}}}}}{{{\beta _{x,l}}\,}}}. 
\end{array}
\end{equation}
The last summation in Eq.~(\ref{EqSm}) was obtained with the aid of Poisson summation. This summation is rapidity converging and van be expressed in closed form using a geometrical series. 
%
This finally leads to an expression for the electric field, with  separation between the first propagating harmonic and the remaining  Floquet harmonics that are all evanescent in the dense grid approximation and as long as the modulation frequency is low enough so that the array can be considered as dense also at the highest temporal harmonic $\omega_n$ that cannot be neglected (see Appendix \ref{sec:Calculating the field in unit cell}),
\begin{widetext}
\begin{equation}\label{E in unit cell}
\vec E\left( {\vec r} \right)\, \approx \,\,\, - \frac{{\hat z\eta k{I_0}}}{{2b}}\left\{ \begin{array}{l}
\frac{{\left( {1 + jy{q_y}} \right)}}{{\sqrt {{k^2} - q_y^2} \,}}\left[ {1 + \frac{{1 - j\left( {a + x} \right)\sqrt {{k^2} - q_y^2}  + j{q_x}a}}{{ja\sqrt {{k^2} - q_y^2}  - j{q_x}a}} + \frac{{1 - j\left( {a - x} \right)\sqrt {{k^2} - q_y^2}  - j{q_x}a}}{{ja\sqrt {{k^2} - q_y^2}  + j{q_x}a}}} \right]\\
 - 2\sum\limits_{l = 1}^\infty  {\frac{{{e^{jy\left( {\frac{{2\pi l}}{b}} \right)}}\left( {1 + j{q_y}y} \right)}}{{\left( {\frac{{2\pi l}}{b} + {q_y}} \right)\,}}\left[ \begin{array}{l}
{e^{ - \left| x \right|\frac{{2\pi l}}{b}}}\left( {1 - {q_y}\left| x \right|} \right) + \frac{{{e^{ - 2\pi l\frac{{a + x}}{b}}}\left( {1 - {q_y}a + j{q_x}a - {q_y}x} \right)}}{{1 - {e^{ - 2\pi l\frac{a}{b}}}\left( {1 - {q_y}a + j{q_x}a} \right)}}\\
 + \frac{{{e^{ - 2\pi l\frac{{a - x}}{b}}}\left( {1 - {q_y}a + j{q_x}a + {q_y}x} \right)}}{{1 - {e^{ - 2\pi l\frac{a}{b}}}\left( {1 - {q_y}a - j{q_x}a} \right)}}
\end{array} \right]}
\end{array} \right\}.
\end{equation}
\end{widetext}

In order to simplify the calculation of the average fields inside the unit cell, since the wire is assumed to be thin $r_0\ll a,b$ we will neglect the fact that the field inside the PEC wire is zero. This simplification enables us to integrate inside the unit cell without complicated boundaries. This assumptions is supported by the fact that, 
\begin{equation}\label{approx in unit cell}
\begin{array}{l}
\frac{{2\pi \int\limits_0^{{r_0}} {{E^w}rdr} }}{{ab}} = 2\pi \frac{{\eta k{I_0}}}{{4ab}}\int\limits_0^{{r_0}} {H_0^{\left( 2 \right)}\left( {{k_0}r} \right)rdr} \\
\,\,\,\,\,\, \sim O\left( {\frac{{r_0^2}}{{ab}}\ln {k_0}{r_0}} \right)
\end{array}
\end{equation}
which implies that the resulting error by this approximation is $O(r_0^2/ab)$.
The average field is then calculated by,
\begin{equation}
\left\langle {\vec E} \right\rangle   \approx \frac{1}{{ab}}\int\limits_{ - \frac{a}{2}}^{\frac{a}{2}} {\int\limits_{ - \frac{b}{2}}^{\frac{b}{2}} {\vec E\left( {\vec r} \right)dxdy} }.
\end{equation}
%
%
%
%
Thus, 
\begin{widetext}
\begin{equation} \label{E_avg}
\left\langle {\vec E} \right\rangle \,\,\, \approx \,\,\, - \frac{{\hat z\eta {k_0}{I_0}}}{2}\left\{ \begin{array}{l}
\frac{2}{{jab\left( {k_0^2 - q_0^2} \right)}} - \frac{1}{{b\sqrt {k_0^2 - q_y^2} \,}}\\
 - 2\sum\limits_{l = 1}^\infty  {\frac{{{q_y}b{{\left( { - 1} \right)}^l}}}{{2l\pi a\left( {\frac{{2\pi l}}{b} + {q_y}} \right)\,}}\left[ \begin{array}{l}
\frac{{ - {q_y}b + 2l\pi  + {e^{ - l\pi \frac{a}{b}}}\left( {{q_y}al\pi  + {q_y}b - 2l\pi } \right)}}{{2{l^2}{\pi ^2}}} + \\
\left( {\frac{{2{e^{ - \left( {2\pi l\frac{a}{b} + {q_y}a - j{q_x}a} \right)}}\sinh \frac{a}{2}\left( {\frac{{2\pi l}}{b} + {q_y}} \right)}}{{b\left( {\frac{{2\pi l}}{b} + {q_y}} \right)}}} \right)\left[ \begin{array}{l}
\frac{1}{{1 - {e^{ - 2\pi l\frac{a}{b}}}\left( {1 - {q_y}a + j{q_x}a} \right)}} + \\
\frac{1}{{1 - {e^{ - 2\pi l\frac{a}{b}}}\left( {1 - {q_y}a - j{q_x}a} \right)}}
\end{array} \right]
\end{array} \right]}
\end{array} \right\}
\end{equation}
\end{widetext}
For a dense array, the infinite summation in Eq.~(\ref{E_avg}) can be neglected, and thus the average field may be simplified as
%
\begin{equation}\label{approx average}
\left\langle {{{\vec E}}} \right\rangle \approx  - \hat z\eta {k_0}{I_0}\left[ {\frac{1}{{jab\left( {k_0^2 - q_0^2} \right)}} - \frac{1}{{2b\sqrt {k_0^2 - q_y^2} \,}}} \right]
\end{equation}
The result in Eq.~(\ref{approx average}) has a good agreement with \cite{Nonlocal permittivity from a quasistatic model for a class of wire media} (see appendix \ref{sec:Heavily dense grid}). Following \cite{Radiation from elementary sources in a uniaxial wire medium}, we can write the relation between the macroscopic (averaged) fields,
\begin{subequations}
\begin{alignat}{3}
&\vec \nabla  \times {{\vec E}_{macro}} =  - j\omega {\mu _0}{{\vec H}_{macro}}\\
&\vec \nabla  \times {{\vec H}_{macro}} = \frac{I}{{ab}}\hat z + j\omega {\varepsilon _0}{{\vec E}_{macro}}.
\end{alignat}
\end{subequations}
Therefore, the macroscopic magnetic field,
\begin{equation}
{{\vec H}_{macro}} =  - \frac{1}{{jk\eta }}\vec \nabla  \times {{\vec E}_{macro}} =  - \frac{1}{{jk\eta }}j{{\vec q}_0} \times \left\langle {\vec E} \right\rangle.
\end{equation}
The average Poynting's vector reads \cite{Radiation from elementary sources in a uniaxial wire medium},
\begin{equation}
{S_{avg}} = \frac{1}{2}{\mathop{\rm Re}\nolimits} \left\{ {{{\vec E}_{macro}} \times {{\vec H}^*}_{macro} + \frac{{{\varphi _w}{I^*}}}{{ab}}\hat z} \right\}.
\end{equation}
And since in our case the currents are independent of $z$, $\varphi_w=0$, and we may write,
\begin{equation}
{S_{avg}} = \frac{1}{2}{\mathop{\rm Re}\nolimits} \left\{ {{{\vec E}_{macro}} \times {{\vec H}^*}_{macro} } \right\} = \frac{{{q_0}}}{{2k\eta }}{\left| {\left\langle E \right\rangle } \right|^2}.
\end{equation}


\section{Perturbed continues media near the plasma region }\label{Sec perturbed cont}
In the previous sections we have homogenized and  derived the effective permittivity of the various modal solutions that are supported by the spatiotemporally modulated wire medium. However, there is another way to go. We could, first homogenize the stationary wire medium. Leading to its famous plasma-like behaviour, and then, introduce the space-time modulation on the effective properties of the homogenized material. Evidently, the latter approach is expected to be less accurate since it does not take into account the microscopic properties of the modulation, however, it is surely easier to perform. In this section we compare the two approaches. Specifically, near its resonance frequency, the effective permittivity of a stationary wire medium takes the plasma-like form,
\begin{equation}\label{eps cont plas}
{\varepsilon _r} = 1 - \frac{{\omega _p^2}}{{{\omega ^2}}}.
\end{equation}
The spatiotemporal modulation of the structure may now be introduced in the effective permittivity parameter $\omega_p$.

In order to be able to perform a fair comparison it is essential to define what it the analogue of the capacitance modulation discussed in the previous sections. Clearly, in the homogenized model, the only quantity we may modulate is $\omega_p$, but then, we need to ask for a given capacitance  modulation $\delta C$ in Eq.~(\ref{capacitors}), what is the equivalent measure of modulation in $\omega_p$?

To that end, we may estimate the behaviour for a small change in $\omega_p$ and the frequency. Recall that in Sec.~\ref{Sec Near Plasma} we have defined the frequency that nullifies the effective permittivity of the loaded, but stationary, wire medium as $\omega^0$ (see Eq.~(\ref{k_0_sqr})). In the continuous model that we consider here in Eq.~(\ref{eps cont plas}), $\omega_p$ takes the role of $\omega^0$. In order to estimate the small variation behaviour of Eq.~(\ref{eps cont plas}) we denote $\omega_p=\omega^0+\delta\omega_p$, and $\omega=\omega^0+\delta\omega$. Then, for frequencies near the plasma frequency, and for small variations in the plasma frequency,  Eq.~(\ref{eps cont plas}) becomes,
\begin{equation} \label{eps_to_freq_and_omega_p}
\begin{array}{l}
{\varepsilon _r} \approx   2\left( {  \frac{{\delta \omega }}{{{\omega ^0}}} - \frac{{\delta {\omega _p}}}{{{\omega ^0}}}} \right).
\end{array}
\end{equation}
By performing a similar procedure  for the effective permittivity of the stationary wire medium that is given in Eq.~(\ref{eps_LTI}), and comparing the permittivity near the plasma frequency, we obtain that a variation in $\omega_p$ is related to a variation in $C$ via,
\begin{equation} \label{delta_omega_p_to_delta_c}
\frac{{\delta {\omega _p}}}{{{\omega ^0}}} =  - \frac{\psi }{2}\,\frac{{\delta C}}{{{C_0}}}
\end{equation}
when $\psi  \ll 1$. If $\psi\gtrapprox 1$ then the simple local plasma model in Eq.~(\ref{eps cont plas}) is inadequate to describe the effective model of the loaded wire medium.
Let us now time-modulate the plasma frequency,
\begin{equation}
{\omega _p}\left( {\vec r,t} \right) = {\omega ^0} + \delta {\omega _p}\cos \left( {\Omega t - \varphi } \right),
\end{equation}
with $\varphi = {\vec \zeta}\cdot {\vec R}$ and ${\vec R} = x\hat{x} + y\hat{y}$. Then, we  approximate,
%
\begin{equation}
\omega _p^2\left( {\vec r,t} \right) \approx {\left( {{\omega ^0}} \right)^2}\left[ {1 + 2\frac{{\delta {\omega _p}}}{{{\omega ^0}}}\cos \left( {\Omega t - \varphi } \right)} \right].
\end{equation}
We assume a $\hat z$ polarized electric field as in the original wire system.  Then, write Faraday's law in the frequency domain $\vec \nabla  \times \vec H\left( \omega  \right) = j\omega \hat z\varepsilon \left( \omega  \right)E\left( \omega  \right)$, and convert to the time domain ($1/j\omega \mapsto$ integration),
\begin{equation}
\vec \nabla  \times \vec H\left( t \right) = \hat z{\varepsilon _0}\frac{{\partial E\left( t \right)}}{{\partial t}} + \hat z{\varepsilon _0}\omega _p^2\left( t \right)\int\limits_{ - \infty }^t {E\left( \tau  \right)d\tau }.
\end{equation}
We can now return back to the frequency domain,
\begin{equation}
\begin{array}{l}
\vec \nabla  \times \vec H\left( \omega  \right) = j\hat z\omega {\varepsilon _0}\vec E\left( \omega  \right) + \\
\,\,\,\,\hat z{\varepsilon _0}{\left( {{\omega ^0}} \right)^2}\left\{ \begin{array}{l}
\delta \left( \omega  \right) + \\
\frac{{\delta {\omega _p}}}{{{\omega ^0}}}\left[ \begin{array}{l}
{e^{j\varphi }}\delta \left( {\omega  - \Omega } \right) + \\
{e^{ - j\varphi }}\delta \left( {\omega  + \Omega } \right)
\end{array} \right]
\end{array} \right\} * \frac{{\vec E\left( \omega  \right)}}{{j\omega }}
\end{array}
\end{equation}
where $*$ denotes convolution.
We will derive the equation for the $n$th frequency harmonics, and also use the known identity for plane waves $ \vec \nabla =- j\vec q  $
\begin{equation}
\begin{array}{l}
 - \frac{1}{{{\varepsilon _0}}}j\vec q \times {{\vec H}_n} = \left[ {{\omega _n} - \frac{{{{\left( {{\omega ^0}} \right)}^2}}}{{{\omega _n}}}} \right]\hat zj{E_n} + \\
\,\,\,\,\,\,\,\,\,\,\,\,\hat z\frac{{{\omega ^0}\delta {\omega _p}{e^{j\varphi }}{E_{n - 1}}}}{{j{\omega _{n - 1}}}} + \,\hat z\frac{{{\omega ^0}\delta {\omega _p}{e^{ - j\varphi }}{E_{n + 1}}}}{{j{\omega _{n + 1}}}}
\end{array}.
\end{equation}
Now using in addition Faraday's equation, we obtain,
\begin{equation}
{q^2}{E_z} =  - j{\omega _n}\mu {\varepsilon _0}\frac{{\left( { - j\vec q} \right)}}{{{\varepsilon _0}}} \times \vec H.
\end{equation}
We will now get an equation for the electric field, which can be written as an infinite tri-diagonal matrix. Let us now focus in the case of linear phase $\varphi  = \vec \zeta  \cdot \vec R$.
\begin{equation} \label{E_n_cont}
a_{n-1}e^{j\varphi}E_{n-1}-b_nE_n + a_{n+1}e^{-j\varphi}E_{n+1}=0,
\end{equation}
with,
\begin{subequations}
\begin{alignat}{3}
&{a_n} =  \frac{{{k_0}}}{{{k_{n}}}}\frac{{\delta {\omega _p}}}{{{\omega ^0}}}\\
&{b_n} = \frac{{{k_n}}}{{{k^0}}} - \frac{{{k^0}}}{{{k_n}}} - \frac{{q_n^2}}{{{k^0}{k_n}}}\\
\end{alignat}
\end{subequations}
By assuming, as in previous sections,  that only the three fundamental harmonics are dominant, the recursive relation in Eq.~(\ref{E_n_cont}) may be truncated. Yielding a $3\times3$ homogenous system for $E_n$, with $n=-1,0,1$. Then, to get a nontrivial solution we require that its determinant vanishes.  Yielding,
\begin{equation} \label{q_0_equation_cont}
\left(\frac{\delta\omega_p}{\omega^0} \right) \left[\frac{a_{-1}}{b_{-1}} + \frac{a_{1}}{b_{1}} \right] = b_0
\end{equation}
in a similar way to which Eq.~(\ref{approximate_determinant_equation}) was derived. Later, we use  Eq.~(\ref{q_n}) and get the dispersion equation for the various wave solutions supported by the system. For the sake of brevity we omit the expressions, nevertheless, these solutions are then used to calculate the effective permittivity of the various modes, as done in Sec.~\ref{Sec Homo}. These read,
%
\begin{widetext}
\begin{subequations}
\begin{align}
&{\varepsilon _{r:1}} = {\varepsilon _{r:2}} = \frac{{\left( {2 + \frac{{\delta \omega }}{{{\omega ^0}}} - \frac{{\delta \hat \omega }}{{{\omega ^0}}}} \right)\left( {\frac{{\delta \omega }}{{{\omega ^0}}} - \frac{{\delta \hat \omega }}{{{\omega ^0}}}} \right)}}{{{{\left( {1 + \frac{{\delta \omega }}{{{\omega ^0}}}} \right)}^2}}}\\
&{\varepsilon _{r:3,4}} = \frac{1}{{{{\left( {1 + \frac{{\delta \omega }}{{{\omega ^0}}}} \right)}^2}}}{\left[ { - \frac{\zeta }{{{k^0}}}\cos \gamma  \pm \frac{{{k_1}}}{{{k^0}}}\sqrt { - {{\left( {\frac{\zeta }{{{k_1}}}} \right)}^2}{{\sin }^2}\gamma  + 1 - \frac{{{k^0}}}{{{k_1}}}} } \right]^2}\\
&{\varepsilon _{r:5,6}} = \frac{1}{{{{\left( {1 + \frac{{\delta \omega }}{{{\omega ^0}}}} \right)}^2}}}{\left[ {\frac{\zeta }{{{k^0}}}\cos \gamma  \pm \frac{{{k_1}}}{{{k^0}}}\sqrt { - {{\left( {\frac{\zeta }{{{k_1}}}} \right)}^2}{{\sin }^2}\gamma  + 1 - \frac{{{k^0}}}{{{k_1}}}} } \right]^2}\\
\end{align}
\end{subequations}
\end{widetext}

\begin{figure}[ptb]
\centering
\vspace{-0.0cm}
\includegraphics[width=\columnwidth]{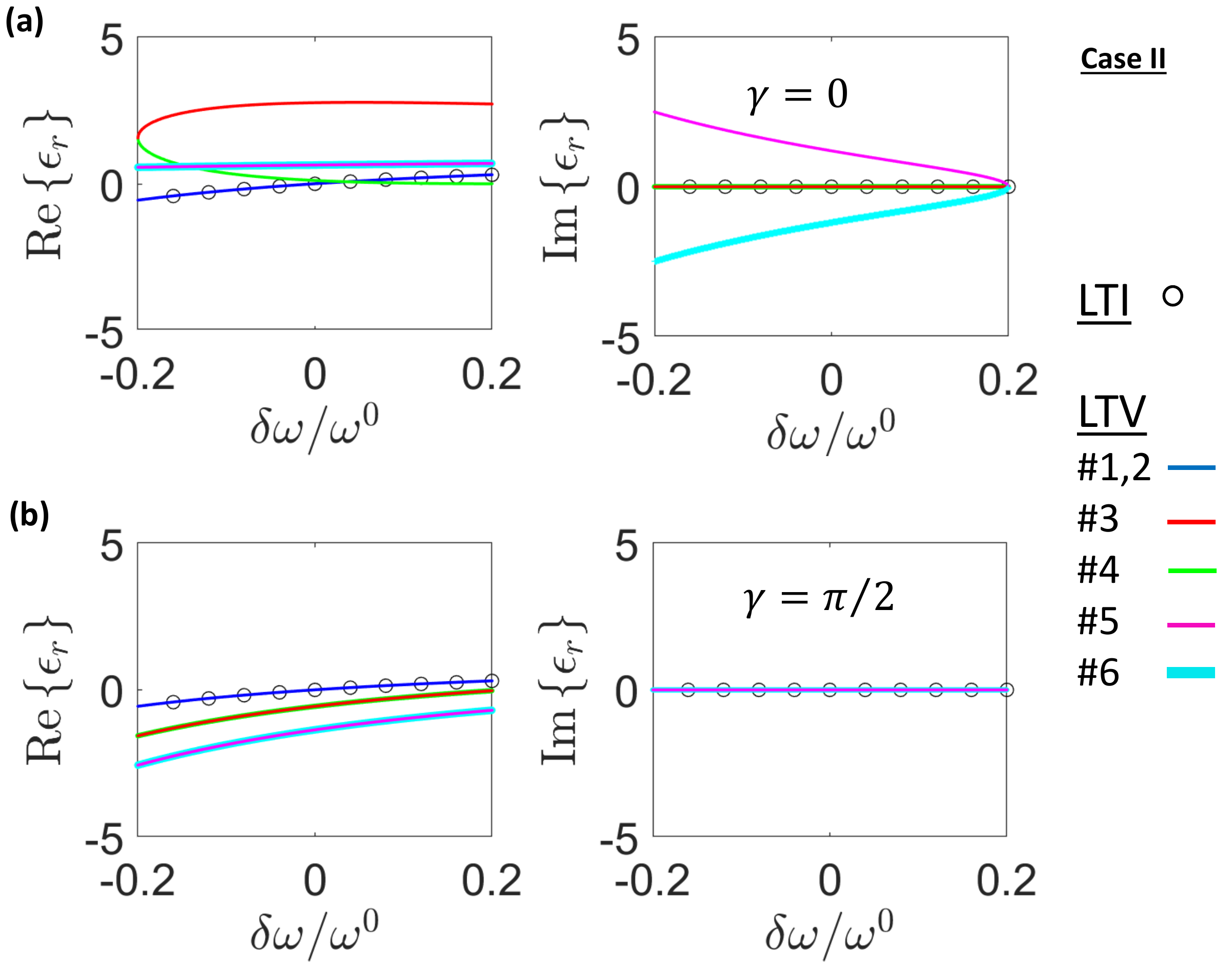}
\vspace{-0.0cm}%
\caption{Effective permittivity of the modal solutions in a spatiotemporally modulated continuous plasma-like model. Results for case II in Table~\ref{Table2}. This figure should be compared with Fig.~\ref{Fig_EffectivePermittivity_wire_case2} that was calculated by brute force homogenization of the spatiotemporally modulated wire medium.
(a) effective permittivity when the propagation is collinear with the modulation. Circles: no modulation - stationary plasma medium. Cont. lines: blue-fundamental modal solutions \#1,2 are nearly identical to these of the stationary medium, and are reciprocal. The other higher order solutions \#3,4 and \#5,6  exhibit  deviation with respect to the stationary medium solutions and nonreciprocally. (b) as (a) but for propagation normal to the direction of modulation. The solutions are reciprocal in this case, and thus the strong anisotropy due to the space-time modulation clearly emerges by comparing the two figures (a) and (b). A comparison with Fig.~\ref{Fig_EffectivePermittivity_wire_case2} demonstrates strong similarities between the results of the two homogenization approaches.}
%
%
\label{Fig_EffectivePermittivity_cont_case2}%
\end{figure}

\begin{figure}[ptb]
\centering
\vspace{-0.0cm}
\includegraphics[width=\columnwidth]{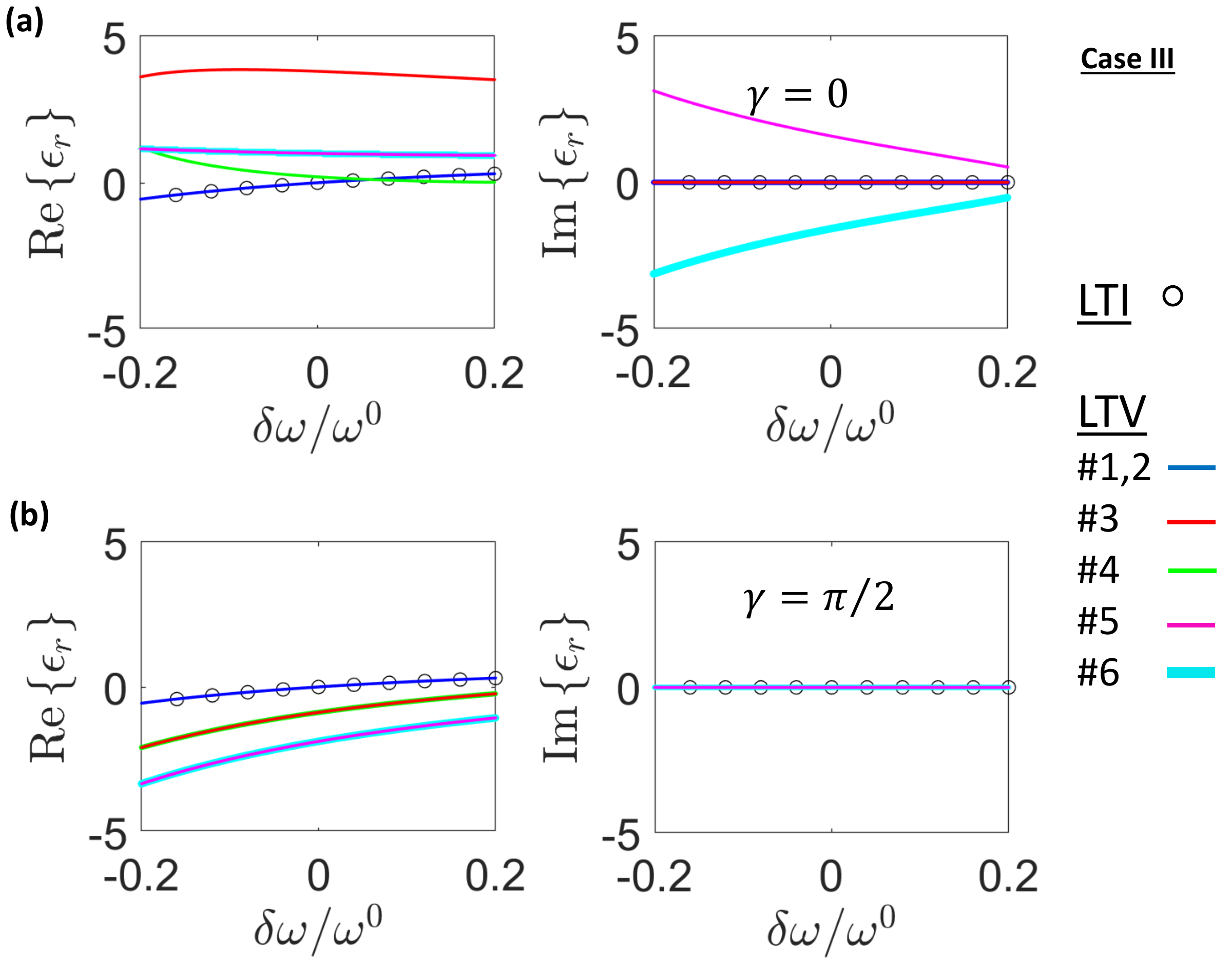}
\vspace{-0.0cm}%
\caption{Effective permittivity, spatiotemporally modulated continuous plasma-like model. As in Fig.~\ref{Fig_EffectivePermittivity_cont_case2} but calculated for case III in Table~\ref{Table2}.   These results should be compared with Fig.~\ref{Fig_EffectivePermittivity_wire_case3}, exhibiting large differences between the two homogenization approaches in the case $\psi\sim1$.}
\label{Fig_EffectivePermittivity_cont_case3}%
\end{figure}

In order to compare with the effective primitivity that is obtained by directly homogenizing the spatiotemporally modulated wire medium we use the transform $\frac{{\delta {\omega_p}}}{{{\omega ^0}}} =  - \frac{\psi }{2}\,\frac{{\delta C}}{{{C_0}}}$ for the cases listed in Table~\ref{Table1}, and thus get a new set of corresponding four cases that are listed in Table~\ref{Table2}. 
\begin{table}
\begin{tabular}{|l|c|c|c|c|c|c|c|}
  \hline
           & $m=\delta\omega_p/\omega^0$    & $\Omega/\omega^0$ & $\zeta/k^0$ & $\theta_0$ & $\xi$  \\ \hline
  Case I   & 0.014  & 0.15              & 0.1         & $\pi/3$    & $0$   \\ \hline
  Case II  & 0.035  & 0.2               & 1.0         & $0$        & $0$ \\ \hline
  Case III & 0.03   & 0.25              & 1.2         & $\pi/5$    & $\pi/4$  \\ \hline
  Case IV  & 0.18   & 0.3               & 1.0         & $\pi$      & $0$  \\
  \hline
\end{tabular}
\caption{Four parameter cases used in the numerical examples for the effective permittivity in a continuous plasma model with spatiotemporally modulated plasma-frequency. The values here correspond to these in Table~\ref{Table1} for the wire medium.}\label{Table2}
\end{table}
where here the modulation depth $m=\delta\omega_0/\omega^0$.
Specifically, we calculate the effective permittivities for the modal solutions that correspond to cases II and III. These are shown in Fig.~\ref{Fig_EffectivePermittivity_cont_case2} and Fig.~\ref{Fig_EffectivePermittivity_cont_case3}, and should be compared with the wire-medium counterparts in  Fig.~\ref{Fig_EffectivePermittivity_wire_case2} and Fig.~\ref{Fig_EffectivePermittivity_wire_case3} above.
Specifically, it is seen that  the two homogenization approaches are nearly equivalent, with minor differences for cases with $\psi\ll1$, as seen for example by comparing Fig.~\ref{Fig_EffectivePermittivity_cont_case2} and Fig.~\ref{Fig_EffectivePermittivity_wire_case2}.
This behaviour changes for cases with larger capacitive loading $\psi\gtrapprox 1$ as seen for example by comparing Fig.~\ref{Fig_EffectivePermittivity_cont_case3} and Fig.~\ref{Fig_EffectivePermittivity_wire_case3}. In this case since the capacitive loading is not negligible, the plasma model in Eq.~(\ref{eps cont plas}) becomes less accurate already for the stationary medium case \cite{Tretyakov}. Nevertheless, while the effect of the spatiotemporal modulation on the fundamental solutions \#1,2 is moderate, the effect on the higher order solutions, that dominate below $\omega^0$, is substantial, despite the fact that the modulation index in this case $m=0.03$ is small.

\section{Conclusion}
In this paper we have developed homogenization theory for spatiotemporally modulated wire medium. Our analysis  takes into account the complete interaction between the time-modulated wires, both in space and time. The dispersion relations that we derive for the modes that are supported by the lattice demonstrate peculiar wave phenomena such as guidance below the cutoff frequency (so called plasma-frequency) of the stationary medium, with modulation depth as weak as desired. This behaviour shares similarities with the extraordinary mode that is guided in magnetized plasma, parallel to the magnetization direction, the so called Whistler mode. Furthermore, in light of the ability to guide below the plasma frequency, we show that this system provides a mean to achieve substantial effective Fresnel drag with weak and slow modulation. In addition to these wave phenomena we calculate, analytically,  the effective permittivity of the first low order solutions, and moreover derive expressions for the averaged (macroscopic) fields, and Poynting's vector. Lastly, we compare between two homogenization approaches of spatiotemporally modulated wire medium, first, that includes the modulation in the homogenization, and second that introduces the modulation into the effective parameters of the homogenized corresponding stationary (LTI) medium. We show that under certain conditions these approaches provide similar results, however, they may substantially deviate in other parameters regimes.

\section*{Acknowledgment}
This research was supported by the Israel Science Foundation (grant No. 1353/19).

\section{Appendixes}
\subsection{Evaluation of the dispersion of LTI wire medium} \label{sec:Evaluate the dispersion of LTI wire media}
This calculation can be found entirely in textbooks, such as in \cite{Tretyakov}, nevertheless, for the sake of completeness and self-continency, we briefly provide it below.  The local field at the reference point $m=l=0$,
\begin{equation}
\begin{array}{l}
E_z^{loc} =  - \frac{{\eta k{I_0}}}{4}\sum\limits_{m,l \ne 0,0} {H_0^{\left( 2 \right)}\left( {k{R_{ml}}} \right){e^{ - j\left( {{q_x}am + {q_y}bl} \right)}}} \\
\,\,\,\,\,\,\,\,\,\,\,\,\, =  - \frac{{\eta k{I_0}}}{2}\left( {{S_0} + \sum\limits_{m \ne 0} {{S_m}{e^{ - j{q_x}am}}} } \right).
\end{array}
\end{equation}
Where $S_m$ denotes the summation for fixed $m$, over the indices $l$. For $m=0$ we have,
\begin{equation}
\begin{array}{l}
{S_0} = \sum\limits_{l = 1}^\infty  {H_0^{\left( 2 \right)}\left( {k{R_{0l}}} \right)\cos {q_y}bl} \\
\,\,\,\,\,\, = \frac{1}{{b\sqrt {{k^2} - q_y^2} }} - \frac{1}{2} + \,\\
\,\,\,\,\,\,\,\,\,\,\,\,\,\,\frac{j}{\pi }\left\{ {\ln \frac{{bk}}{{4\pi }} + \gamma  + \frac{1}{2}\sum\limits_{l \ne 0} {\left[ {\frac{{ - 2\pi j}}{{b{\beta _{x,l}}}} - \frac{1}{{\left| l \right|}}} \right]} } \right\}.
\end{array}
\end{equation}
Here $\gamma  = 0.5772$ is the Euler constant.
For $m\neq0$ we use the Poisson summation which becomes highly efficient for dense grid, and get,
%
\begin{equation}
\begin{array}{l}
{S_m} = \frac{1}{2}\sum\limits_{l =  - \infty }^\infty  {H_0^{\left( 2 \right)}\left( {k{R_{ml}}} \right){e^{ - j{q_y}bl}}} \\
\,\,\,\,\,\,\,\, =  - \frac{1}{b}\sum\limits_{l =  - \infty }^\infty  {\frac{{{e^{ - ja\left| m \right|{\beta _{x,l}}}}}}{{{\beta _{x,l}}\,}}}.
\end{array}
\end{equation}
By substituting Eq.~(\ref{S_m_media}) to Eq.~(\ref{E_local_media}), and changing the order of summations, and using,
\begin{equation} \label{S_m}
\sum\limits_{m \ne 0} {{e^{ - j\left( {{\beta _{x,l}}a\left| m \right| + {q_x}am} \right)}}}  = \frac{{j\sin {\beta _{x,l}}a}}{{\cos {\beta _{x,l}}a - \cos {q_x}a}} - 1.
\end{equation}
We can rewrite Eq.~(\ref{dispersion_cond}),
\begin{equation}
\begin{array}{l}
\frac{{\eta k}}{4}H_0^{\left( 2 \right)}\left( {k{r_0}} \right) + \frac{1}{{j\omega \tilde{C}_0 }}\, = \\
\,\,\,\,\,\, - \frac{{\eta k}}{2}\left\{ {{S_0} + \sum\limits_{l =  - \infty }^\infty  {\frac{1}{{{\beta _{x,l}}}}\left[ {\frac{{j\sin {\beta _{x,l}}a}}{{\cos {\beta _{x,l}}a - \cos {q_x}a}} - 1} \right]} } \right\}
\end{array}
\end{equation}
Note that we assumed small losses in he background medium, so that convergence is secured. We will assume that the wire radius is very small comparing to the separation between the wires, and use the approximation for the Hankel function \cite{Bessel function},
\begin{equation} \label{H_approx}
H_0^{\left( 2 \right)}\left( {k{r_0}} \right) \approx 1 - j\frac{2}{\pi }\left( {\ln \frac{{k{r_0}}}{2} + \gamma } \right).
\end{equation}
This leads to the dispersion relation,
\begin{equation}
\begin{array}{l}
\frac{j}{\pi }\ln \frac{{k{r_0}}}{2} - \frac{2}{{jk\eta \omega \tilde{C}_0 }}\, = \frac{1}{{b\sqrt {{k^2} - q_y^2} }} + \\
\,\,\,\,\,\,\frac{j}{\pi }\left[ {\ln \frac{{bk}}{{4\pi }} + \frac{1}{2}\sum\limits_{l \ne 0} {\left( {\frac{{ - 2\pi j}}{{b{\beta _{x,l}}}} - \frac{1}{{\left| l \right|}}} \right)} } \right] + \\
\,\,\,\,\,\sum\limits_{l =  - \infty }^\infty  {\frac{1}{{{\beta _{x,l}}}}\left[ {\frac{{j\sin {\beta _{x,l}}a}}{{\cos {\beta _{x,l}}a - \cos {q_x}a}} - 1} \right]}.
\end{array}
\end{equation}
And for a dense grid ($ka,kb \ll 1$), with the approximations for trigonometric functions with small arguments we can get,
\begin{equation}
\begin{array}{l}
\ln \frac{b}{{2\pi {r_0}}} + \sum\limits_{l = 1}^\infty  {\frac{{\coth \frac{{\pi al}}{b} - 1}}{l}}  - \frac{{2\pi }}{{\eta v{k^2}\tilde{C}_0 }}\\
\,\,\,\,\,\,\,\, = \frac{{2\pi }}{{ab\left[ {{k^2} - \left( {q_x^2 + q_y^2} \right)} \right]}} - \frac{{\pi a}}{{6b}}.
\end{array}
\end{equation}

\subsection{Derivation of Eq.~(\ref{E_time})} \label{sec:susceptibility of time variant loaded wire}
\begin{figure}
{ \centering
 \includegraphics[width=\columnwidth]{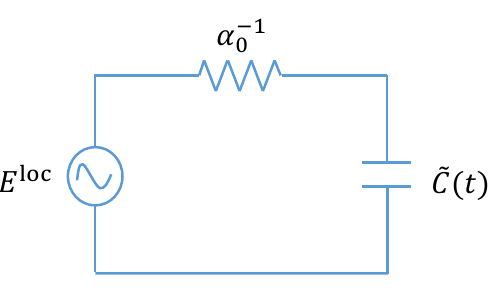}\\
 \caption{The equivalent circuit of a wire illuminated by an external field}\label{equivalent circuit}
}
\end{figure}

We begin by transforming the expression for the frequency domain susceptibility of a loaded wire, i.e., Eq.~(\ref{gamma_of_loaded_wire}), into the time-domain. It easy to notice that $\alpha^{-1}$ behaves as an impedance to unit length, and $E^{loc}$ behaves as an  external ``voltage'' source (obviously, with dimensions  [V/m]). Therefore, we can consider our physical model to be equivalent to the simple serial circuit that is shown in Fig.~\ref{equivalent circuit} with stationary impedance  $\alpha_0^{-1}$ per unit length (involves resistance as well as reactance) and time modulated capacitance.
The current on the capacitor, reads
\begin{equation}
I = \frac{d}{{dt}}\left( {\tilde C\left( t \right){E_C}\left( t \right)} \right)
\end{equation}
where $E_C$ denotes the ``voltage'' on the capacitance. Then, Eq.~(\ref{E_time}) is straightforward.
%

\subsection{Derivation of Eq.~(\ref{q_n})} \label{sec:Dispersion of the harmonics}
The wire media behaves as an infinite periodic crystal, and therefore we can use Floquet-Bloch theorem, with the expression in Eq.~(\ref{I_m_l_with_perturbation}). By substituting Eq.~(\ref{I_m_l_with_perturbation}) into Eq.~(\ref{E_inc_to_I_n}), and using the explicit expression for the  phase of the capacitors ${\varphi _{m,l}} = \vec \zeta  \cdot \vec r = ma\zeta \cos \xi  + bl\zeta \sin \xi$. We get for the $m,l$ wire, and for n-th mode:
\begin{equation} \label{E_inc_to_I_n_every_wire}
\begin{array}{l}
E_n^{inc}{e^{ - j{{\vec q}_n} \cdot \vec{R}_{ml}}} =
\left[ {{\gamma _0} {{\omega _n}  }  + \frac{1}{{ j {{\omega _n} } {\tilde{C}_0}}}} \right]{A_n}{e^{ - j{{\vec q}_n} \cdot \vec{R}_{ml}}} - \\
A_{n-1}\frac{m}{2}\frac{1}{j\omega_{n-1}\tilde{C}_0}e^{-j(\vec{\zeta} + \vec{q}_{n-1})\cdot\vec{R}_{ml}}-
\\
A_{n+1}\frac{m}{2}\frac{1}{j\omega_{n+1}\tilde{C}_0}e^{-j(-\vec{\zeta} + \vec{q}_{n+1})\cdot\vec{R}_{ml}}.
\end{array}
\end{equation}
The expression in (\ref{E_inc_to_I_n_every_wire}) must be valid for all the wires $m,l$. Therefore, we must require that
%
%
%
\begin{equation}
 - \vec \zeta  - {{\vec q}_{n - 1}} =  - {{\vec q}_n},\mbox{ and }
\vec \zeta  - {{\vec q}_{n + 1}} =  - {{\vec q}_n}
\end{equation}
Therefore, we get ${{\vec q}_n} = {{\vec q}_0} + n\vec \zeta $ as in Eq.~(\ref{q_n}).

\subsection{Calculating the field in unit cell} \label{sec:Calculating the field in unit cell}
In order to evaluate the electric field in the unit cell at the origin, we use Eq.~(\ref{E_unit_cell_origin}),  
\begin{equation}
\begin{array}{l}
\vec E\left( {\vec r} \right) = \\
\,\,\,\, - \frac{{\hat z\eta k{I_0}}}{{2b}}\sum\limits_{m =  - \infty }^\infty  {\sum\limits_{l =  - \infty }^\infty  {\frac{{{e^{jy\left( {{q_y} + \frac{{2\pi l}}{b}} \right)}}{e^{ - ja\left| {m - \frac{x}{a}} \right|{\beta _{x,l}}}}{e^{ - j{q_x}am}}}}{{{\beta _{x,l}}\,}}} }
\end{array}
\end{equation}
Then we approximate ${\beta _{x,l}} =  - j\sqrt {{{\left( {{q_y} + \frac{{2\pi l}}{b}} \right)}^2} - {k^2}}  \approx  - j\left( {\frac{{2\pi l}}{b} + {q_y}} \right)$. This leads to,
\begin{equation}
\begin{array}{l}
\vec E\left( {\vec r} \right) =  - \frac{{\hat z\eta k{I_0}}}{{2b}} \cdot \\
\,\,\,\,\,\,\sum\limits_{l =  - \infty }^\infty  {\frac{{{e^{jy\left( {{q_y} + \frac{{2\pi l}}{b}} \right)}}}}{{{\beta _{x,l}}}}\left[ \begin{array}{l}
\frac{{{e^{ - j\left( {a + x} \right){\beta _{x,l}}}}{e^{j{q_x}a}}}}{{1 - {e^{ - ja{\beta _{x,l}}}}{e^{j{q_x}a}}}} + {e^{ - j\left| x \right|{\beta _{x,l}}}}\\
 + \frac{{{e^{ - j\left( {a - x} \right){\beta _{x,l}}}}{e^{ - j{q_x}a}}}}{{1 - {e^{ - ja{\beta _{x,l}}}}{e^{ - j{q_x}a}}}}
\end{array} \right]}
\end{array}
\end{equation}
For a dense grid  ${\beta _{x,l}}$ may be further approximated,
\begin{equation}
{\beta _{x,l}} \approx \left\{ {\begin{array}{*{20}{c}}
{\sqrt {{k^2} - q_y^2} }&{l = 0}\\
{ - j\left( {\frac{{2\pi \left| l \right|}}{b} + {q_y}} \right)}&{l \ne 0}
\end{array}} \right.
\end{equation}
We can see that we get ${\beta _{x,l}} \approx {\beta _{x, - l}}$.
%
These eventually lead to Eq.~(\ref{E in unit cell}). 

\subsection{Heavily dense grid} \label{sec:Heavily dense grid}

We compare Eq.~(\ref{approx average}) with the average over the $z$-component of the electric field as calculated  in \cite{Nonlocal permittivity from a quasistatic model for a class of wire media},
\begin{equation} \label{maslovski_avg_field}
\left\langle { E_z} \right\rangle  = \left( {j\omega \tilde L + {Z_w}} \right)I + \frac{{\partial \varphi }}{{\partial z}}
\end{equation}
Where ${\tilde L}$ is the effective inductance per unit length of the wire, and ${Z_w}$ is the impedance loading on the wire. In our case, for a PEC wire with capacitance load, we have ${Z_w} = \frac{1}{{j\omega \tilde C_0}}$. Also, since we are focusing on the case of a propagating wave in a direction normal to the wires, there is no variation along the $z$-axis and thus we can nullify the potential derivative $\frac{{\partial \varphi }}{{\partial z}} = 0$. Then we get  (\ref{maslovski_avg_field}),
\begin{equation} \label{maslovski_avg_field_sub}
\begin{array}{l}
\left\langle {{E_z}} \right\rangle  =  jI\left[ {\left( {1 + \frac{{\delta \omega }}{{{\omega ^0}}}} \right){\omega ^0}\tilde L - \frac{1}{{\left( {1 + \frac{{\delta \omega }}{{{\omega ^0}}}} \right){\omega ^0}\tilde C_0}}} \right]\\
\,\,\,\,\,\,\,\,\, \approx jI\frac{\eta }{{ab{k^0}}}\left[ {1 + \left( {1 + 2\psi } \right)\left( {\frac{{\delta \omega }}{{{\omega ^0}}}} \right)} \right]
\end{array}
\end{equation}
The expression in (\ref{maslovski_avg_field_sub}) is valid for a very dense grid in a stationary medium. Now to show that our averaged field in Eq.~(\ref{approx average}) is consistent with that. Note that the second term in Eq.~(\ref{approx average}) should be neglected for a heavily dense grid. Then,  using the Eq.~(\ref{q_0_1}) for the wave number in the media (with zero perturbation $\delta C = 0$), we get, 
\begin{equation}
\begin{array}{l}
\left\langle {\vec E} \right\rangle \, \approx  - \frac{{\hat z\eta {k_0}{I_0}}}{{ab}}\left\{ {\frac{1}{{j\left( {k_0^2 - q_0^2} \right)}}} \right\}\\
\, \approx  - \frac{{\hat z\eta {k_0}{I_0}}}{{ab}}\left\{ {\frac{1}{{j\left( {k_0^2 - {{\left( {{k^0}} \right)}^2}{{\left( {1 + \frac{{\delta \omega }}{{{\omega ^0}}}} \right)}^2}\left[ {1 - \frac{1}{{{{\left( {1 + \frac{{\delta \omega }}{{{\omega ^0}}}} \right)}^2}\left( {1 + \psi } \right) - \psi }}} \right]} \right)}}} \right\}\\
\, \approx j\frac{{\hat z\eta {I_0}}}{{ab{k^0}}}\left\{ {\frac{{1 + \frac{{\delta \omega }}{{{\omega ^0}}}}}{{1 + 2\frac{{\delta \omega }}{{{\omega ^0}}} - 2\left( {1 + \psi } \right)\frac{{\delta \omega }}{{{\omega ^0}}}}}} \right\}\\
\, \approx j\frac{{\hat z\eta {I_0}}}{{ab{k^0}}}\left\{ {1 + \left( {1 + 2\psi } \right)\frac{{\delta \omega }}{{{\omega ^0}}}} \right\}.
\end{array}
\end{equation}
Identical to Eq.~(\ref{maslovski_avg_field_sub}).


\end{document}